# Supporting Real-Time COVID-19 Medical Management Decisions: The Transition Matrix Model Approach

Jian Chen Ph.D., CreditWise Technology Co. Ltd.
Michael C. Fu, Ph.D., University of Maryland
Zhang Wenhong[1], M.D., Huashan Hospital, Fudan University
Zheng Junhua[2], M.D., Shanghai First People's Hospital, Shanghai Jiaotong University

June 30, 2020

## Abstract

Since the onset of the COVID-19 outbreak in Wuhan, China, numerous forecasting models have been proposed to project the trajectory of coronavirus infection cases. Most of these forecasts are based on traditional epidemiological models. However, many of these forecasts have performed poorly, mainly due to two reasons: 1) these type of forecast models are highly sensitive to model parameters, which have wide confidence intervals; 2) the models fail to incorporate the non-pharmaceutical-intervention (or NPI, mainly government policies, like "lock-down", "shut-down", "stay-at-home" directives) effects successfully. We propose a new discrete-time Markov chain model that directly incorporates stochastic behavior and for which parameter estimation is straightforward from available data. Transition matrix models (TMM) have been widely used in financial industry, mainly in credit analysis, especially in predicting credit rating migration of corporate bonds, or delinquency migration of consumer loans. The event chain of a consumer loan's "early delinquency", "serious delinquency", "default", is very like the event chain of COVID-19's "mild case", "severe case", "critical case", "death".

Using such data from China's Hubei province (for which Wuhan is the provincial capital city and which accounted for approximately 82% of the total reported COVID-19 cases in the entire country), the model is shown to be flexible, robust, and accurate. As a result, it has been adopted by the first Shanghai assistance medical team in Wuhan's Jinyintan Hospital, which was the first designated hospital to take COVID-19 patients in the world. The forecast has been used for preparing medical staff, intensive care unit (ICU) beds, ventilators, and other critical care medical resources and for supporting real-time medical management decisions.

---

[1] Dr. Zhang is the head of the Center for Infectious Disease at Huashan Hospital of Fudan University and serves as the leader of Shanghai Anti-COVID-19 clinical expert team.

[2] Dr. Zheng is the vice president of First People Hospital, which is affiliated with Shanghai Jiaotong University. As the leader of the first Shanghai medical assistance team to Wuhan after the coronavirus outbreak beginning in January, he and his team worked on the frontlines battling the pandemic for 67 days and nights.

Empirical data from China's first two months (January/February) of fighting COVID-19 was collected and used to enhance the model by embedding NPI efficiency into the model. We applied the model to forecast Italy, South Korea, and Iran on March 9. Later we made forecasts for Spain, Germany, France, US on March 24. Again, the model has performed very well, proven to be flexible, robust, and accurate for most of these countries/regions outside China.

Compared to widely used SIR-type models, we find TMM more flexible, robust, and accurate for COVID-19 forecasts. More importantly, it has been adopted by frontline medical professionals to support real-time COVID-19 medical management decisions and proven to be a more pragmatic forecasting tool. Three out of the four authors were invited to provide insights on this novel model at the Brookings event "Fighting COVID-19: Experiences and lessons from the frontlines in Asia"[3], which was broadcasted international news agencies, including Caixin Global[4], China Daily[5], Reuters, etc. The model has also received wide attention in news media, including Caixin, with a full feature article on its forecasts[6].

We suggest that modeling teams around the world should take a closer look at this discrete-time Markov chain model and examine its usefulness in the battle against COVID-19, including the support for real-time decision making of preparing medical staff, equipment, and other medical resources, as well as government/business planning of when and how to impose/lift non-pharmaceutical intervention (NPI) policies.

# 1. Introduction

Novel coronavirus pneumonia (COVID-19) caused by zoonotic 2019 novel coronavirus (SARS-Cov-2) broke out in Wuhan, China during December 2019, and was declared a pandemic by WHO on March 11, 2020. As of June 30, 2020, more than 10 million laboratory-confirmed cases and 500,000 deaths have been documented worldwide. Among the tools used in the fight to contain such pandemics, forecast models are critical in helping support not only medical resource management decisions, but also in informing government policies, such as when and where "lock-down" and "stay-at-home" directives should be enacted and lifted.

Traditional epidemiology models use deterministic differential equations to forecast the population dynamics among various states, e.g., susceptible, exposed, infectious, or recovered in the well-known susceptible-exposed-infectious-recovered (SEIR) model recently used to model the COVID-19 outbreak in Wuhan (Wu, Leung, and Leung 2020). Such models have also been expanded to include additional states for COVID-19, e.g., Giordano et al. (2020). These models are by design aggregate models that track only the mean populations and do not incorporate stochastic effects directly and are highly sensitive to estimated parameters. For example, a key parameter is the basic reproductive number $R_0$, or the rate (average number) at which one currently infected person infects new persons; thus, an $R_0 < 1$ indicates that the epidemic is dying out. An inaccurate estimate of $R_0$ is magnified in poor forecasts, leading to orders of magnitude differences in output.

---

[3] https://www.brookings.edu/events/webcast-only-fighting-covid-19-experiences-and-lessons-from-the-frontlines-in-asia/
[4] https://www.caixinglobal.com/2020-04-02/fighting-covid-19-experiences-and-lessons-from-the-frontlines-in-asia-101537685.html
[5] https://www.chinadaily.com.cn/a/202004/04/WS5e87f516a310128217284656.html
[6] http://international.caixin.com/2020-03-14/101528645.html

More recently, epidemiologists have been looking at agent-based models, very familiar to the operations research (OR) community, where individuals can be modeled in detail, e.g., age, gender, health condition, and stochastic characteristics are directly incorporated. Such models can be very useful for studying smaller communities, but since they typically require stochastic (Monte Carlo) simulation, may face computational challenges in scaling up to large cities or countries when the number of agents becomes large. Also, if the amount of detailed individual data available is limited, then it would be challenging to estimate the model with any degree of confidence.

We propose a discrete-time Markov chain (DTMC) model, where the states of the chain are similar to the states in the compartmental models. Since this DTMC modeling approach is analogous to a financial forecasting model called the transition matrix model (TMM), widely used in credit analysis (e.g., Malik and Thomas 2012, Chen et. al 2018), be it corporate rating migration, or individual consumer behavior, we will also refer to it as the TMM approach. For example, most major rating agencies, such as Morningstar[7], Moody's Investor Services[8], S&P Global Rating[9], all publish their annual transition matrices for corporate rating migration, so corporate bond investors can estimate the likelihood of their investments getting downgraded in the future.

Transition matrices can also be stochastic, where the transition probabilities are dependent on both the individual characteristics and other external variables. For example, IFE Group (2012-2015[10]) performed actuarial studies for the world's largest monoline insurance program, Federal Housing Administration (FHA)'s Mutual Mortgage Insurance (MMI) fund, with approximately $2.7 trillion in assets under management. They provided forecasts for the fund's performance for as long as 37 years into the future. And tens of millions of mortgages are simulated for hundreds of paths, to calculate the future expected credit losses, and loss distribution. The calculation process is very time consuming, and can take as long as three hours, even after simulation optimization. Transition matrices can also be hybrid (partially deterministic, partially stochastic), where some probabilities are constant, and some are time varying.

Transition matrix models share similarities with the above-mentioned agent-based model used in epidemiology study. Our proposed approach has the following advantages: it incorporates stochastic features directly while retaining essentially the same states (compartments) as in the compartmental models; its discrete-time nature and degree of modeling detail make it straightforward to estimate model parameters from available data; it is computationally tractable both in terms of parameter estimation and in terms of model output analysis.

---

[7] https://ratingagency.morningstar.com/PublicDocs/Exhibit%201.pdf
[8] https://www.moodys.com/sites/products/ProductAttachments/Credit%20Risk%20Calculator.pdf
[9] https://www.spratings.com/documents/20184/774196/2016+Annual+Global+Corporate+Default+Study+And+Rating+Transitions.pdf/2ddcf9dd-3b82-4151-9dab-8e3fc70a7035
[10] https://www.hud.gov/program_offices/housing/rmra/oe/rpts/actr/actrmenu

The TMM approach has been shown to be flexible, robust, and accurate. From field experience, the TMM forecast results have been used to support real-time medical management decisions, allocate medical staff, and plan for business re-opening. Here are some examples of the TMM forecasts used to support decision making in the fight against COVID-19:

- On February 10$^{th}$, the TMM forecast was adopted by the first Shanghai medical assistance team (led by Dr. Zheng, one of this paper's authors) in Wuhan's Jinyintan Hospital, the first designated hospital to take COVID-19 patients in the world. The forecast has been used in preparing medical staff, ICU beds, ventilators, and other critical care medical resources by central and provincial health commissions and local CDCs.
- On February 14$^{th}$, we published an article[11], indicating that under the cautiously optimal scenario, medical staff needed for taking care of severe and critical patients could reach 40,000-45,000. Soon after this forecast, more medical assistance teams were dispatched from all over China to Wuhan and other cities in Hubei province. The total number of medical assistance teams reached 346, with more than 42,600 medical staff, on March 8[12].
- We also published our forecast for Italy, South Korea, and Iran[13] on March 9$^{th}$, and channeled the Italy forecast to one Italian cabinet member on the same day, indicating a very dire situation with a forecast of more than 190,000 cases likely to occur by April end. The Italian government implemented a national "lock-down" policy on the next day.
- On February 15th, we forecasted the "back-to-normal" date for Hubei province was most likely to be mid-April (4/13-4/20)[14]. On April 8$^{th}$, the "lock-down" in Wuhan was lifted, and Dr. Zheng was able to come back to Shanghai after fighting COVID-19 for 67 days in the epicenter of Wuhan.

This paper is organized as follows. In section 2, we briefly discuss SIR-type models, and discuss their major drawbacks. In section 3, we present the transition matrix model for COVID-19 forecast and draw the analogy to transition matrix models used in the mortgage finance industry. In section 4, we discuss model robustness, accuracy, and flexibility of TMM, compared with SIR-type models. In section 5, we discuss the effects of medical assistance teams dispatched from all over China to Wuhan and other cities in Hubei province, which effectively increased the cure rate and reduced the fatality rate, which is the major reason for the forecast errors for the cure and death tolls.

---

[11] http://chenjian.blog.caixin.com/archives/221560
[12] https://news.sina.com.cn/c/2020-03-08/doc-iimxyqvz8817389.shtml
[13] http://chenjian.blog.caixin.com/archives/223401
[14] http://chenjian.blog.caixin.com/archives/221630

## 2. Specification of SEIR/SIR-Type Models

Prevailing epidemiological forecast models relevant to COVID-19 are based on extensions of the susceptible-exposed-infectious-recovered (SEIR) model, which are deterministic continuous-time dynamic models that model the evolution of the aggregate population under consideration, where the population is separated into a fixed number of mutually exclusive "compartments". For example, in the original susceptible-infectious-recovered (SIR) model of Kermack and McKendrick (1927), the compartments are defined as follows:

- Susceptible (S) – not infected yet;
- Infected (I) – assumed infectious with symptoms;
- Removed (R) – recovered (sometimes this is the definition) or deceased.

The three compartments are represented by time-varying functions S(t), I(t), R(t), representing the (average) number in each compartment (state). The simplest set of ordinary differential equations (ODEs) modeling the dynamics is the following:

$$\frac{dS(t)}{dt} = -aR_0 \frac{I(t)S(t)}{N},$$

$$\frac{dI(t)}{dt} = a(R_0 \frac{I(t)S(t)}{N} - I(t)),$$

$$\frac{dR(t)}{dt} = aI(t),$$

which has just three parameters (a>0, $R_0$>0, and the population N). This system can be solved analytically, but once any realistic features are incorporated into the model, which is the case for real-world applications, numerical simulation is required, e.g., the SEIR model of Wu, Leung, and Leung (2020) used for COVID-19.

Generally, compartmental models can be very effective epidemiology tools once a disease is well understood and in a mature phase. They are also good theoretical models for reference purposes, e.g., for comparing the infection rate of COVID-19 against other respiratory infectious diseases such as SARS and MERS, by comparing the different values of $R_0$, once accurate estimates of model parameters can be obtained. However, for forecasting purposes based on relatively sparse data, especially with regards to patient-level outcomes (as opposed to aggregate cases), they may exhibit high sensitivity to the estimated model parameters such as $R_0$, limiting their robustness in forecasting in the early and middle stages of epidemics. In the SEIR model of Wu, Leung, and Leung (2020), $R_0$ is estimated "using Markov Chain Monte Carlo methods with Gibbs sampling and non-informative flat prior," so as we will shortly see, it is a more involved process than the parameter estimation process for our proposed approach, which relies directly on available empirical data.

There are several serious drawbacks to SIR-type models. The first drawback is its high sensitivity to model parameters, which poses serious challenges in terms of model robustness. A basic model driver in compartmental models (SIR, SEIR, and their extensions) is the parameter $R_0$, which indicates how contagious an infectious disease is, and is also referred to as the reproduction number, because it represents the average number of people who will contract the disease from one person with that disease.

Compartmental models are very sensitive to their model parameters, as well as initial conditions, so accurate estimates are crucial if these models are to be used for forecasting, which is not necessarily the primary usage for these types of models, especially in the early stages of a new outbreak. A wide range of $R_0$ values were reported by different researchers, ranging from 2.0 to 5.7, as listed in Table 1 below. Given this level of difference, the forecasted number of infection cases will differ by 200% in one week, 800% in two weeks, and 8000% in one-month's time! If after 5 million infection cases and more than 330,000 deaths (May 21 data), scientists still cannot agree on the very fundamental parameter, that probably suggests there is no magic $R_0$ that is universally applicable. And forecasts based on estimated $R_0$ will be highly unstable. Table 1 summarizes the forecast numbers of cumulative cases for the $R_0$ estimates, assuming incubation period of 5.2 days, as estimated in Li et al. (2020). Assuming the initial condition is 100 infection cases, within three weeks, forecasts vary from as low as 224 to as high as 3.5 million! Even for the same paper, the forecast numbers can vary widely, and the max to min forecast ratio ranges from 5.3 to 720. This table clearly illustrates high parameter sensitivity exhibited by SIR-type models.

Table 1: $R_0$ Estimates in Various Papers

| Paper | $R_0$ Estimate | 95% CI | Total Infection in 3 Weeks from 100 Infection Cases | | | |
|---|---|---|---|---|---|---|
| | | | #Cases Min | #Cases Exp | #Cases Max | Max/Min Ratio |
| Majumder and Mandl, 2020 | 2.65 | 2.0 - 3.3 | 485 | 1,123 | 2,597 | 5.3 |
| Imai et al., 2020 | 2.6 | 1.5–3.5 | 255 | 1,053 | 3,361 | 13.2 |
| Liu et al., 2020 | 2.92 | 2.28–3.67 | 697 | 1,590 | 4,185 | 6.0 |
| Riou and Althaus, 2020 | 2.2 | 90% CI: 1.4–3.8 | 224 | 628 | 4,949 | 22.1 |
| Li et al., 2020 | 2.2 | 1.4–3.9 | 224 | 628 | 5,630 | 25.1 |
| Imperial College COVID-19 Response Team | 3.87 | 3.01–4.66 | 1,786 | 5,416 | 15,006 | 8.4 |
| Sanche et. al. 2020 | 5.7 | 3.8 - 8.9 | 4,949 | 57,397 | 3,560,865 | 719.5 |

In one SIR-extended model (Giordano, Blanchini, Bruno, et al. 2020), the forecasts for Italy's COVID-19 outcome are 40%, 0.25%, and 0.09% of total population infected, thus the max/min ratio is close to 444 times, which basically make the forecast impractical for medical resource planning.

Secondly, SIR-type models cannot easily incorporate NPI measures. They generally assume that reduction of transportation and social interaction will reduce the coronavirus transmission gradually, which is reflected in the reduction of inflow and outflow of passengers, and transmissibility parameters such as $R_0$. However, drastic NPI measures taken by governments, such as lock-down of city blocks, communities, and mass quarantine of close contacts not only affect the inflow and outflow of people but also reduce the infection period, as close contacts are quickly tracked and quarantined. These features are not well captured by such models. As a result, some models over-forecast the infection cases by a large factor, e.g., Wu et al. (2020a) estimated there would be 800,000 COVID-19 cases in Shanghai by the end of February. But due to massive testing and contact tracing efforts of more than 700 public health officials and 300,000 volunteers, all the close contacts of imported infection cases were quickly identified and quarantined. As a result, there was not a single case of community transmission and only 337 cases were reported by February end, for a metropolis of approximately 30 million people.

In summary, from a practical perspective, if the goal of the forecasting model is to provide support for decisions such as medical resource planning, including staff (doctors, nurses, cleaning staff, etc.), supplies (ICU beds, ventilators, N95 masks, etc.), and biohazardous waste disposal capacity, alternative modeling approaches may be more aligned with the available data for estimating model parameters. Furthermore, adding new compartments in SIR/SEIR models to allow differentiation between severe cases (requiring hospital beds) and critical cases (requiring ICU beds and ventilators) results in additional model complexity and the introduction of even more parameters that are difficult to estimate from the available data.

## 3. Specification of Transition Matrix Model

We model the patient treatment process as an absorbing Markov chain with the following discrete states (analogous to compartments in the SIR/SEIR traditional differential equation-based models): (under) medical observation, discharged, infected non-severe, infected severe, critical, death, and cured. The potential transitions between the states are shown in Figure 1, where self-loops are understood but omitted in the diagram for clarity, and the three states outside the treatment boxed labelled discharged, cured, and dead are absorbing states. Note that "Infected" (represented as a decision diamond) is not a separate state by itself in the DTMC model, as once a close contact is a confirmed case of infection, it is immediately classified as severe or non-severe. Another state called infected asymptomatic could also be easily added to the model, but since there is sparse data to estimate this state[15], we have not included it, and those patients would not have entered the medical observation state in the current version of the model.

One of the main differences between the compartment models and the proposed modified DTMC model is that the former relies on parameters that have interpretative meanings, so specific data are required to estimate them, whereas the parameters of the proposed model depend only on transition probabilities between states defined based on medical classifications where available data can be used to estimate them directly. As an example, in the SEIR model of Wu, Leung, and Leung (2020), there are two parameters corresponding to the mean latent and infectious periods, which are challenging to estimate accurately from early data.

---

[15] Back in February, asymptomatic cases were not reported in China.

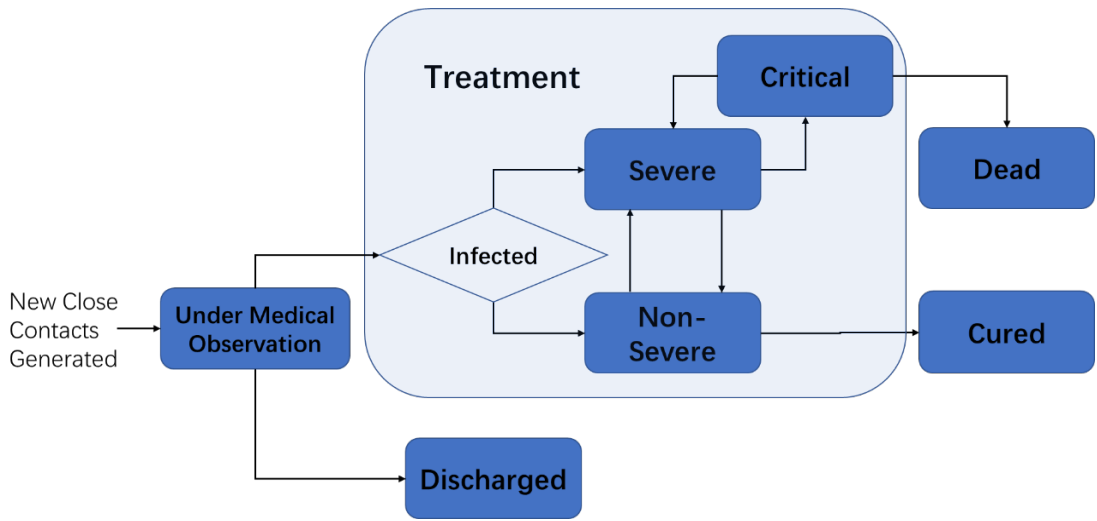

Figure 1. Patient State Transition Diagram

Note that the states "Cured" and "Discharged" could also have been put together into a single state (e.g., called "Healthy"), but for purposes of parameter estimation and tracking statistics, the cure rate is of separate interest. The three states, Discharged, Deceased, and Cured, are absorbing states, whereas the others are all transient states, assuming non-zero probabilities for existing transitions in the diagram.

This is very similar to the transition matrix model (TMM) widely used in mortgage finance industry. In the following chart, we can see a normal (current) mortgage can go to early delinquency for missing 1-2 payments. If the borrower can make up the payments, then she can return to the normal state. If she keeps missing payments, her mortgage will be moved to the serious delinquency (SDQ) department for special treatment. Sometimes even the mortgage servicing will be transferred to a special servicer who is more sophisticated with SDQ loan servicing. One option is to go through a loan modification via different payment reduction schemes, i.e., term extension, rate reduction, principal forgiveness and/or forbearance. If a borrower can successfully go through loan modification, she may return to the normal state. If a modification is not working out, the loan may go to default.

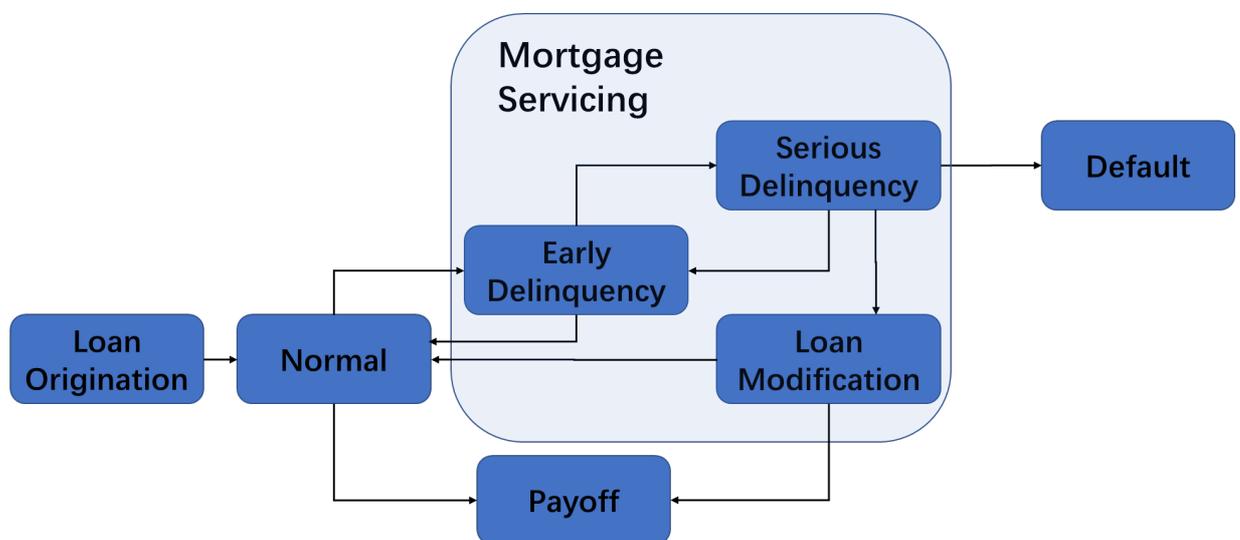

Figure 2. Mortgage State Transition Diagram

The structures of Figures 1 and 2 share many similarities, which convinced us to try a TMM approach to predict the COVID-19 progression. The transition probabilities of the DTMC model can be either determined by regression or simply derived from empirical probabilities.

Table 2 compares SIR-type models and the proposed TMM approach.

Table 2. Comparison of SIR-Type Model and TMM Approach

| SIR vs. TMM | | SIR-type Models | TMM Approach |
|---|---|---|---|
| Characteristics | Model Feature | Deterministic | Stochastic |
| | Temporal Feature | Continuous Time | Discrete Time |
| | Functional Form | Dynamic PDEs | Transition Probability |
| Parameters & Error Distribution | # of Parameters | High | Low |
| | Confidence Interval | Relatively Wide | Determined by Empirical Data |
| | Parameter Sensitivity | Very High | Relatively Low |
| | Error Distribution | Hard to get, mainly determined by C.I. | Generated by Simulation |
| Flexibility | Time-Varying Parameters | Not Allowed | Allowed |
| | NPI Inclusion | Hard to incorporate | Easy to incorporate with empirical probabilities |
| | New State Inclusion | Hard to add new state without introducing more parameters | Easy to incorporate with empirical probabilities |

The first application of TMM was to predict the COVID-19 progression in China's Hubei province, for which Wuhan is the provincial capital city and which accounted for approximately 82% of the total reported COVID-19 cases in the entire country. We adopted the empirical approach when estimating the model in our application to predicting the COVID-19 spread in Hubei province, China. Specifically, we defined the states of the model as follows:
- Under Medical Observation (UMO): state of a close contact of a potential infection case, who is traced, identified and put into medical observation, generally in a quarantine facility. From this state, the next day a patient may be confirmed with infection, discharged without infection, or remain UMO (e.g., if the test results have not come back yet).

- Discharged (Dis): terminal state for a close contact after undergoing medical observation, although one could possibly reenter as a close contact in the generation phase.
- Infected but Non-Severe (INS): state for a patient with mild symptoms (generally treated in makeshift shelter hospitals for Hubei COVID-19 patients). There are three possible states for the next day: cured, worsened condition to severe, or remaining non-severe.
- Infected and Severe (IAS): state for a patient who develops severe symptoms that require hospitalization and oxygen support (WHO 2020). There are again three possible states for the next day: worsening to critical, improvement to non-severe, or staying the same.
- Infected and Critical (IAC): state for a patient showing critical symptoms and requiring admission to an ICU (WHO 2020). There are also three possible states for the next day: death, improvement to severe, staying the same.
- Cured (Cu) and Deceased (De) are two terminal states.

For severe and critical cases, COVID-19 can be complicated by the acute respiratory distress syndrome (ARDS), sepsis and septic shock, multiorgan failure, including acute kidney injury and cardiac injury. Treatment for these patients is very complicated and requires extremely skilled specialists. For example, one of this paper's authors, Dr. Zheng led the first Shanghai Medical Assistance Team to Wuhan, with 150 medical specialists. They worked at the ICU in the world's first designated COVID-19 hospital, Jinyintan Hospital, for 67 days, treating 170 severe and critical patients, with a cure rate of more than 80%. Among these patients, 123 were critical cases, with a cure rate of 72.35%.

Note that "infected" itself is not an independent state, since the outcome is revealed instantaneously. In addition to these states, there are a few other COVID-19 numbers worth mentioning, and we also provide forecasts for these numbers:
- Daily new confirmed cases (DNCC)
- Cumulative confirmed cases (CCC)
- Daily active cases, or infected and under treatment (IAT)

Thus, each day, a patient's state on day $n$ is characterized by a state vector, defined as follows:

$$V(n) = [UMO \quad Dis \quad INS \quad IAS \quad IAC \quad Cu \quad De]',$$

where the value of each element can be viewed as the probability of being in that state, so for each individual new close contact generated, the elements would sum to 1. For a given individual in a known state, the corresponding element would be equal to 1, and the other elements would be zero. For example, a patient currently in state "infected and non-severe" (INS) would have state vector [0 0 1 0 0 0 0]', and the next day, the state vector could remain the same or transition to either [0 0 0 1 0 0 0]' (severe) or [0 0 0 0 0 1 0]' (cured). Thus, for the entire potentially exposed population, the state vector is defined as the count of people in each state. For example, at an early stage with say 100 patients being tested or treated, out of which 30 are awaiting testing results, 10 are critical, 10 are severe, and 50 are non-severe, the state vector would be as follows: [30 0 50 10 10 0 0]'.

Defining the one-step (i.e., daily) transition matrix $P=[p_{i,j}]$, where

$$p_{i,j} = daily\ transition\ probability\ from\ state\ i\ to\ state\ j,$$

we have the usual DTMC vector-matrix (one-step and multi-step) dynamic equations:

$$V(n + 1) = PV(n),$$
$$V(n + m) = P^m V(n),$$

since the count of a certain state comes from itself, all other possible transitions into the state (e.g., INS has two possible incoming states, UMO and IAS), minus the outcome states (IAS, and Cu).

If the population is limited and the transition matrix is stationary, the above formula will be sufficient in predicting all future outcomes, and of course, since it is an absorbing chain, all the transient states would eventually go to zero, and individuals already in the pool will eventually end up discharged, cured, or deceased (the absorbing states). In epidemic settings such as the COVID-19 situation in early 2020, the population is not fixed, and additional individuals enter into the population via the generation of new close contacts in Figure 1. Every day, new close contacts are added to the medical observation pool. In our modified DTMC model, these new individuals are generated as a proportion of the current new close contacts, i.e., via

$$NCC(n + 1) = NCC(n)\ e^{NCC\_Change\_Rate},$$

where the NCC change rate parameter is analogous to the basic reproductive number $R_0$ parameter in SIR-type models, in that when it is positive, the number of infected individuals in the population is increasing, corresponding to an $R_0$ value greater than 1. Just as when intervention measures cause $R_0$ to eventually decrease below 1, correspondingly the NCC change rate will become negative. The NCC change rate is a critical parameter in our forecast, but it is readily estimated from available data, because it appears in a single equation, whereas $R_0$ appears simultaneously in two (or more) equations in SIR and SEIR models (e.g., the very simplest SIR model in Section 2 and the SEIR model of Wu, Leung, and Leung 2020).

This DTMC state transition matrix model can be used for forecasts such as when the infection peak time (maximum number of active infection cases) occurs, as well as patient distributions (critical, severe, non-severe), which can be used for supporting medical resource allocation planning.

Although there are three hospitalization states – non-severe cases (INS), severe cases (IAS), and critical cases (IAC), for the COVID-19 Hubei province data, we had only the patient count in each state but not the actual pairwise transitions among these three states, so we combined the entire hospitalization period into a therapeutic state "infected and being treated" (IAT) – also known as the daily active cases, which minimizes the need for estimation for those unobserved transitions. Instead, we used the proportion of patients of each state (INS, IAS, IAC) within IAT, to forecast the number of non-severe, severe, critical patients. In addition, the model also tracks daily new confirmed cases (DNCC) and cumulative confirmed cases (CCC), since these metrics are tracked in actual data. From all these various data, the parameters of the model can be estimated directly as follows:

- New Close Contacts (NCC) Change Rate : $\alpha_{NCC} = \ln(NCC(t)/NCC(t-1))$;
- UMO Daily Discharge Rate: $p_{UMO,Dis} = Dis(t)/UMO(t-1)$;
- Pr {UMO → IAT}: $p_{UMO,IAT} = IAT(t)/UMO(t-1)$;
- Pr {IAT → Deceased}: $p_{IAT,De} = De(t) / IAT(t-1)$;
- Pr {IAT → Cured}: $p_{IAT,Cu} = Cu(t) / IAT(t-1)$.

And as alluded to above, estimating the count of active severe cases and critical cases requires two additional parameters:

- proportion of severe patients: $\rho_s = IAS(t)/ IAT(t)$;
- proportion of critical patients: $\rho_c = IAC(t)/ IAT(t)$.

We collected all data from Caixin Data (a subsidiary of Caixin Group), who retrieves the original data from China National Health Commission. We also collected supplementary data from Hubei Health Commission, mainly for IAS, IAC count. The data period starts from 2019/12/31 and ends on 2020/2/8, updated on a daily basis. Next, we describe how we choose parameters based on these observed empirical probabilities. The following empirical transition probabilities were observed on 2020/2/8.

Table 3. Empirical Probabilities Observed on Feb. 8th

| Empirical Prob. Observed On 2/8 | 2020/2/8 | 10-Day Moving Average | 10-Day Max | 10-Day Min |
|---|---|---|---|---|
| NCC Change Rate | −21.4% | 3.8% | 71.6% | −21.4% |
| UMO Discharge Rate | 16.5% | 9.0% | 16.5% | 4.1% |
| Pr {UMO → IAT} | 3.04% | 4.54% | 5.39% | 3.04% |
| Pr {IAT → Deceased} | 0.35% | 0.63% | 0.97% | 0.35% |
| Pr {IAT → Cured} | 1.40% | 0.91% | 1.44% | 0.60% |
| Severe Case Ratio | 16.5% | 14.2% | 18.1% | 11.5% |
| Critical Case Ratio | 4.65% | 4.13% | 5.3% | 4.1% |

As mentioned in the previous section when describing the DTMC model, the NCC change rate ($\alpha_{NCC}$) is closely linked to $R_0$ but can be directly estimated from empirical data. This parameter has very strong policy implications, as it measures the effectiveness and efficiency of non-pharmaceutical intervention (NPI) policies and actions. A positive value indicates that the NPI measures are failing, as new close contacts are increasing, whereas a negative value indicates that the NPI measures are effective, as fewer people are contracting the coronavirus on a daily basis. The higher the absolute value of the parameter, the more effective the NPI measures are. Since the 8-day empirical average of -6% is volatile, we considered three values for the model: -1%, -5%, -10%. We could not obtain the one-day probability of UMO discharge rate in Hubei Province, so we used the national rates. Since the one-day probability was 17%, and the 10-day moving weighted average was 13%, we tested the model with both 17% and 13%. In Hubei Province on 2/8, the one-day probability of transition from medical observation to confirmed infection was 2.15%, with a 10-day weighted average of 3.94%, so we used 4% as our model parameter. The latest single-day fatality rate was 0.35% and the 10-day moving weighted average was 0.63%. As this probability was declining, we used the latest value (0.35%) as our model parameter to be cautiously optimistic. The latest one-day cure rate was 1.40%, with a 10-day moving weighted average of 0.91%. As this probability was increasing, we used the latest value (1.40%) as our model parameter. From the historical data, the proportion of critical cases is relatively stable, whereas the proportion of severe cases fluctuated more. Neither of these showed monotone behavior, so we used the average (14.50%, and 4.50%) of the latest value and 10-day moving weighted average as our model parameters to put more weight on recent observations.

In addition to the parameter values just described, we considered six different scenarios to control for forecast uncertainty, based on three different values of the NCC Change Rate $\alpha_{NCC}$ (-10% optimistic, -5% cautiously optimistic, and -1% relatively pessimistic) and two different values of the UMO discharge rate $p_{UMO,Dis}$ (17% and 10.5%, where 17% is more optimistic.). In Wu, Zheng, and Chen (2020), NCC Change Rate was assumed constant, whereas in follow-up work (Zheng et al. 2020), it was allowed to be time-varying to capture changes due to NPI implementations. Table 4 summarizes the scenarios and parameter values used in our model.

Table 4. Parameter Values (in %) Used in the Model Forecast for Different Scenarios.

| Scenario Description | S1: Optimistic | | S2: Cautiously Optimistic | | S3: Relatively Pessimistic | |
|---|---|---|---|---|---|---|
| NCC Change Rate | −10.0% | −10.0% | −5.0% | −5.0% | −1.0% | −1.0% |
| $\alpha_{NCC}$ | 17% | 10.5% | 17% | 10.5% | 17% | 10.5% |
| $p_{UMO,Dis}$ | 4.00% | 4.00% | 4.00% | 4.00% | 4.00% | 4.00% |
| $p_{UMO,IAT}$ | 0.35% | 0.35% | 0.35% | 0.35% | 0.35% | 0.35% |
| $p_{IAT,De}$ | 1.40% | 1.40% | 1.40% | 1.40% | 1.40% | 1.40% |
| $p_{IAT,Cu}$ | 14.50% | 14.50% | 14.50% | 14.50% | 14.50% | 14.50% |
| $\rho_s$ | 4.50% | 4.50% | 4.50% | 4.50% | 4.50% | 4.50% |
| $\rho_c$ | | | | | | |

A summary of the daily update equations used in the model are as follows:

$NCC(t)=NCC(t-1)\exp(\alpha_{NCC})$;
$DNCC(t)=UMO(t-1)p_{UMO,IAT}$;
$CCC(t)=CCC(t-1)+DNCC(t)$;
$UMO(t)=UMO(t-1)+NCC(t)-DNCC(t)-Dis(t)$;
$Dis(t)=UMO(t-1)p_{UMO,Dis}$;
$IAT(t)=IAT(t-1)+DNCC(t)-IAT(t-1)p_{IAT,De}-IAT(t-1)p_{IAT,Cu}$;
$IAS(t)=IAT(t)\rho_s$;
$IAC(t)=IAT(t)\rho_c$;
$INS(t)=IAT(t)-IAS(t)-IAC(t)$;
$De(t)=De(t-1)+IAT(t-1)p_{IAT,De}$;
$Cu(t)=Cu(t-1)+IAT(t-1)p_{IAT,Cu}$.

The primary differences among these scenarios are the different values of NCC Change Rate. **This parameter has a very strong policy implication**, as it measures the effectiveness and efficiency of non-pharmaceutical intervention (NPI) policies and actions. A parameter value greater than zero indicates the NPI measures have failed, and new close contacts will keep increasing. If the NPI measures are effective, this value should be less than zero, and less people contract the coronavirus on a daily basis. The higher the absolute value of the parameter, the more efficient the NPI measures are. The first generation TMM (Wu, Zheng, Chen 2020) used constant NCC Change Rate, whereas the second generation TMM (Zheng et al. 2020) used a time-varying approach to capture different NPI effectiveness and efficiency of different countries more realistically.

## 4. Model Accuracy, Robustness, and Flexibility

In this section, we discuss the model accuracy via back-testing in Section 4.1. Then we discuss model robustness, with respect to parameter sensitivity in Section 4.2. Afterwards we discuss model robustness, with respect to an external shock event in Section 4.3. Lastly, we discuss model flexibility in terms of adjusting to preventive policy effectiveness in Section 4.4.

## 4.1 Model Accuracy

After February, we performed back-testing for the forecast with the actual numbers. Key performance metrics, such as peak value of critical cases, active cases, month end total cases were very close to model forecast under the "cautiously optimistic" scenario, which was picked by us as the most likely scenario. The relative errors of peak active cases, peak severe cases, peak critical cases, and February-end total cases were 1.0%, 20.1%, 7.5%, 1.3%, relative to the median of the two cautiously optimistic scenarios.

Table 5. Back-testing of the Hubei Province Forecast

| Key Performance Metrics | Actual | Forecast | | | | | |
|---|---|---|---|---|---|---|---|
| | | S1: Optimistic | | S2: Cautiously Optimistic | | S3: Relatively Pessimistic | |
| NCC Change Rate | −9.0% | −10% | −10% | −5% | −5% | −1% | −1% |
| UMO Discharge Rate | 16.0% | 17.0% | 10.50% | 17.0% | 10.50% | 17.0% | 10.50% |
| Peak NAT | 50,633 | 39,612 | 47,148 | 44,082 | 55,150 | 62,041 | 85,502 |
| Peak Date | 2020/2/16 | 2020/2/23 | 2020/2/28 | 2020/3/1 | 2020/3/7 | 2020/4/6 | 2020/4/14 |
| Peak NAS | 9,289 | 5,753 | 6,845 | 6,400 | 8,004 | 9,000 | 12,402 |
| Peak Date | 2020/2/16 | 2020/2/23 | 2020/2/28 | 2020/3/1 | 2020/3/7 | 2020/4/6 | 2020/4/14 |
| Peak NAC | 2,492 | 1,786 | 2,124 | 1,986 | 2,484 | 2,793 | 3,849 |
| Peak Case | 2020/2/21 | 2020/2/23 | 2020/2/28 | 2020/3/1 | 2020/3/7 | 2020/4/6 | 2020/4/14 |
| CCC on 02/29 | 66,907 | 54,189 | 64,064 | 60,192 | 71,596 | 68,045 | 81,284 |

Regarding forecast accuracy, SIR-type models generally do not perform very well. As mentioned earlier, Giordano, Blanchini, Bruno, et al. (2020) gave three forecast scenarios for Italy's COVID-19 total infection case estimates while the medium case of 0.25% of total population infected is the closest to the actual observation. The actual infection case on 4/30 was 0.34% of total population infected, so the closest forecast was off by about 36% at the end of April. In Zheng, Wu, Yao, et al. (2020), the TMM forecast for Italy's total infected cases on 4/30 was 192,593, while the actual was 203,591, with an 6% error. Regarding the peak of daily new cases, the TMM forecast was 6,681 on 3/23, while the actual peak occurred on 3/22, with 6,557 cases, or a 2% error. The forecast and back-testing are provided in Appendix I.

## 4.2  Model Sensitivity to Major Parameters

We performed similar parameter sensitivity practice for the TMM used for the Hubei forecast in Wu et al. (2020b). Two major parameters are the NCC change rate, and medical release rate, where NCC change rate takes values of -10%, -5%, and -1%, and medical release rate takes values of -17% and -10.5%.

As can be seen from the following table, while the major parameter NCC_Change_Rate changes from -10% to -1% (a ten-fold difference), and the UMO Discharge Rate change from 17% to 10.5%, the max/min ratio is merely 1.5. This suggests the TMM approach is more robust than SIR-type models.

Table 6. TMM Model Robustness

| Transition Matrix Model | | | | 3 Weeks from Forecast Date | | | |
|---|---|---|---|---|---|---|---|
| Paper | NCC Change Rate | | | #Total Infection Cases | | | Max/Min Ratio |
| Wu, Zheng, Chen(2020) Hubei Forecast Model | S1 | S2 | S3 | S1 | S2 | S3 | |
| UMO Discharge Rate=17% | -10% | -5% | -1% | 54,189 | 60,192 | 68,045 | 1.5 |
| UMO Discharge Rate=10.5% | -10% | -5% | -1% | 64,064 | 71,596 | 81,284 | |

## 4.3 Model Robustness to External Shocks

Next, we take a more detailed look by considering the daily dynamics through the end of February. On February 12[th], three days after we published our forecast (Wu, Zheng, and Chen 2020), the Hubei Province health commission changed the diagnosis criteria, and allowed those patients who do not have definitive PCR test results but have clinical symptoms (mainly CT scan results) to be counted as confirmed COVID-19 cases. This changed criterion increased the daily incremental cases by more than 14,000, resulting in the spike shown in Figure 3.

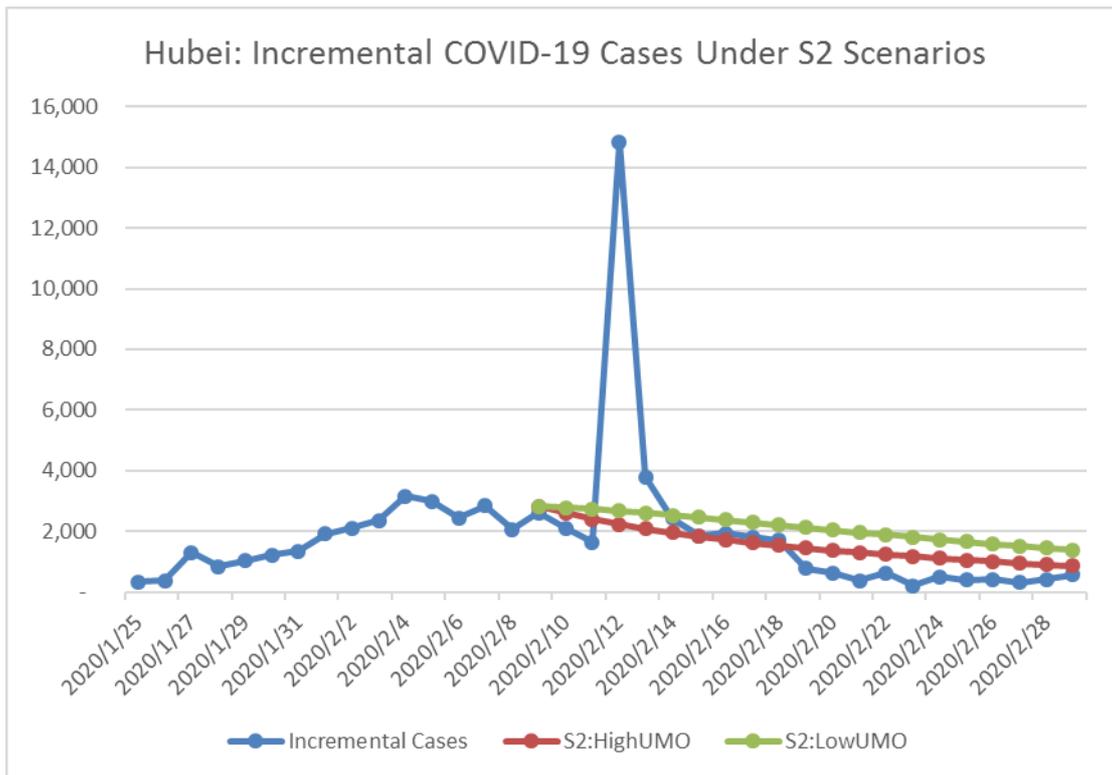

Figure 3. Cautiously Optimistic Scenario Forecasts of Incremental Cases

The spike led us to consider a drastic revision of our model's future forecasts to adjust for the changed criterion. However, after carefully examining the implication of the new criteria, we decided not to change our forecast, as we concluded the change would not have too much impact on the forecast results for cumulative cases in the long term, because it just confirmed suspected cases earlier, most of which would eventually turn into confirmed cases. As long as our forecasts for new close contacts are not heavily affected, the model should be robust to handle this surge in incremental COVID-19 cases. Indeed, Figure 4 confirms that the incremental close contacts did not change dramatically as a result of the spike in new confirmed cases. As a result, the cumulative cases are back in the range of the two S2 scenarios by the end of February, as shown in Figure 5. Also, the critical case numbers are not heavily impacted by the spike, as there is an intermediate state of severe cases between critical cases and new cases; thus, the spike in new cases is not immediately reflected in the critical case numbers, so Figure 6 shows the model forecasts up to February 22 are good, after which the results of the arrival of medical assistance teams led to a dramatic decrease. On the other hand, the severe cases forecasts are clearly impacted by the surge, since there is no middle state between the severe state and new cases, and that resulted in moderate underprediction for the much of the month, as shown in Figure 7.

The arrival of armies of medical assistance teams throughout the Hubei province in February clearly had a beneficial effect on reducing mortality rates and conversely increasing cure rates in Hubei province, shown in Figures 8 and 9, respectively, where the model forecasts gradually and consistently diverge for both of these cumulative counts (cured and deceased) throughout the month. This is discussed further in the next section. As a result, the numbers of cumulative cases are back in the range of S2 scenarios at the end of February.

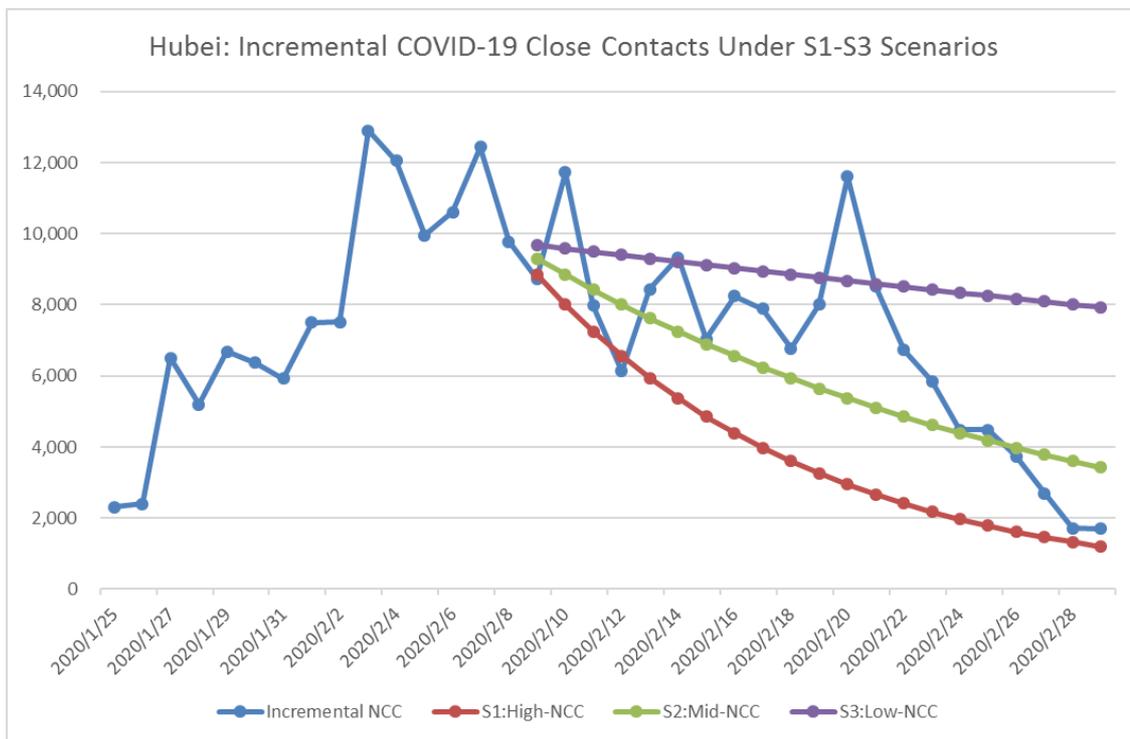

Figure 4. Scenario Forecasts of Close Contacts

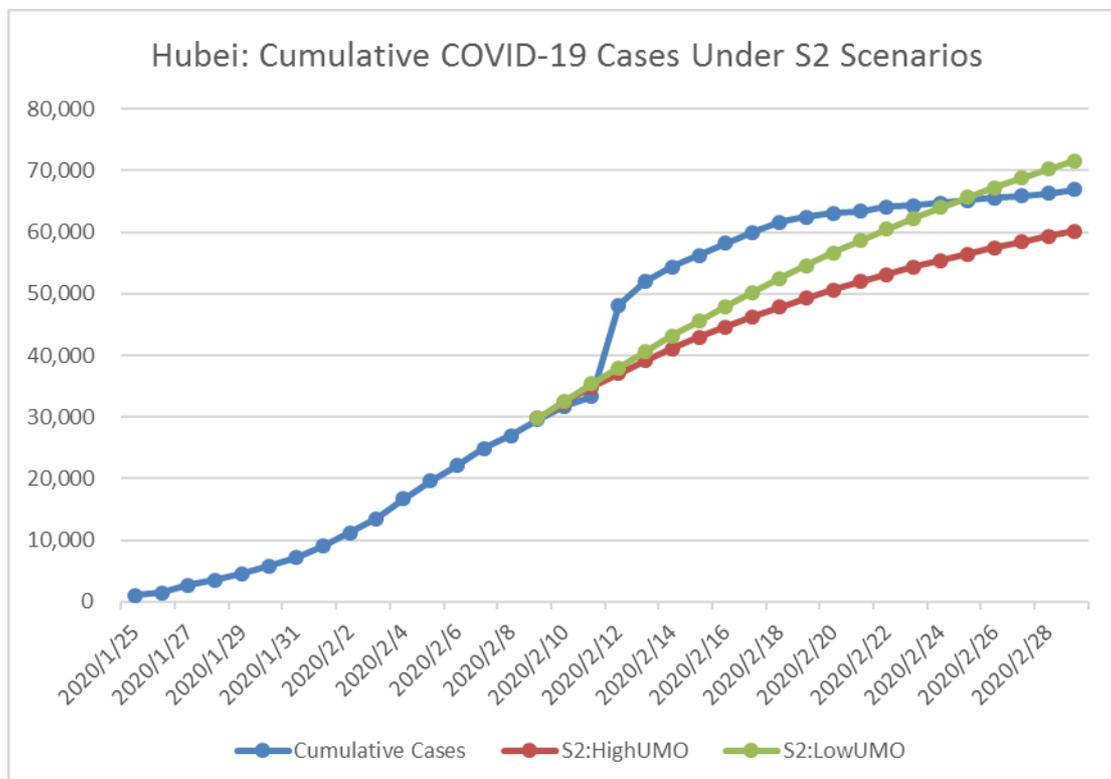

Figure 5. Cautiously Optimistic Scenario Forecasts of Cumulative Cases: The effects of the spike in reported cases on February 12 are washed out by the end of the month.

Also, the number of critical cases is not heavily impacted by the surge, as there is an intermediate state of severe cases between critical cases and new cases, so the surge in new cases is not immediately reflected in the number of critical cases, and our model performed very well before February 22, as seen in Figure 6.

The severe cases forecast is impacted by the surge, since there is no middle state between the severe state and new case, and that resulted in moderate underprediction for the about 10 days, as seen in Figure 7.

The death toll forecast was also close to actual, until late February, as indicated in Figure 8. The improvement in February was mainly driven by the efforts of medical assistance teams dispatched from other provinces to Hubei, and will be described in more detail in Section 5. The cure forecast probably has the biggest forecast error, seen in Figure 9. This is also due to dramatic improvements in medical resources and care, closely related to the previous explanation for the decrease in death rate, so this will also be described in Section 5.

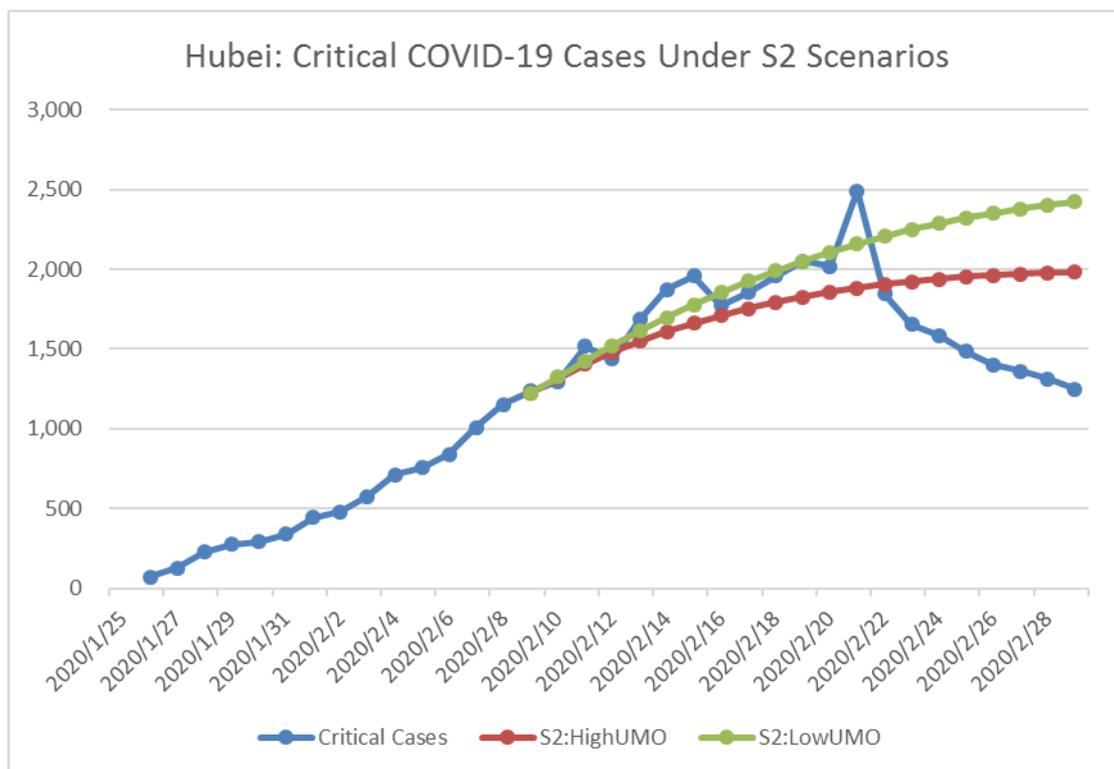

Figure 6. Cautiously Optimistic Scenario Forecasts of Critical Cases: Drop Starting February 22nd Reflecting the Arrival of Medical Assistance Teams from the Rest of China

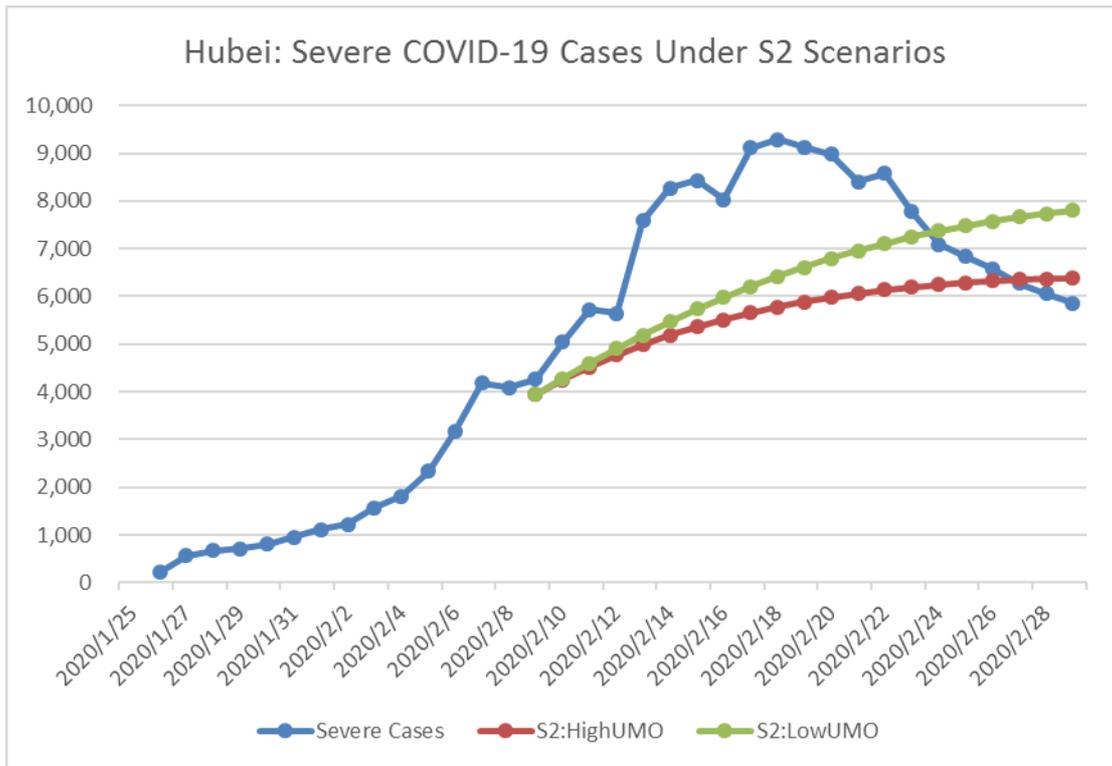

Figure 7. Cautiously Optimistic Scenario Forecasts of Severe Cases: Surge in Reported Cases Due to Changed Criteria (Spike in Figure 2 on February 12) Leads to Underprediction for Most of the Month

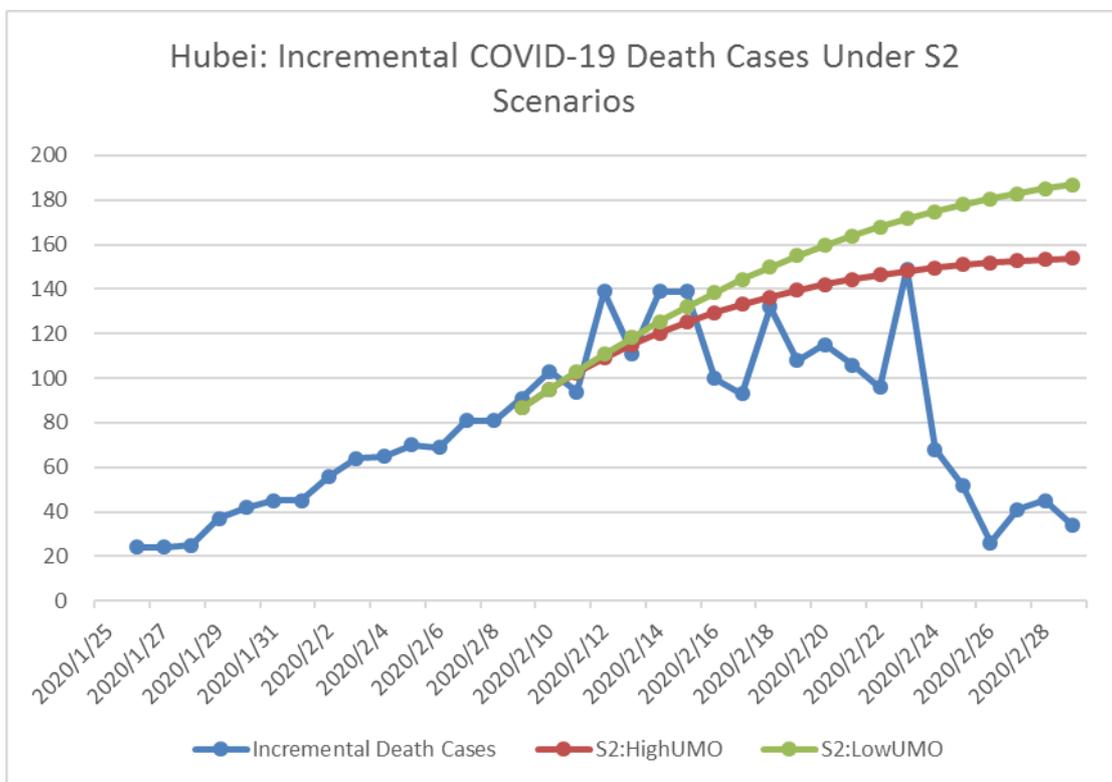

Figure 8. Cautiously Optimistic Scenario Forecasts of Death Cases: Effect of Medical

Assistance Teams Leads to Marked Decreased Mortality Rates During February

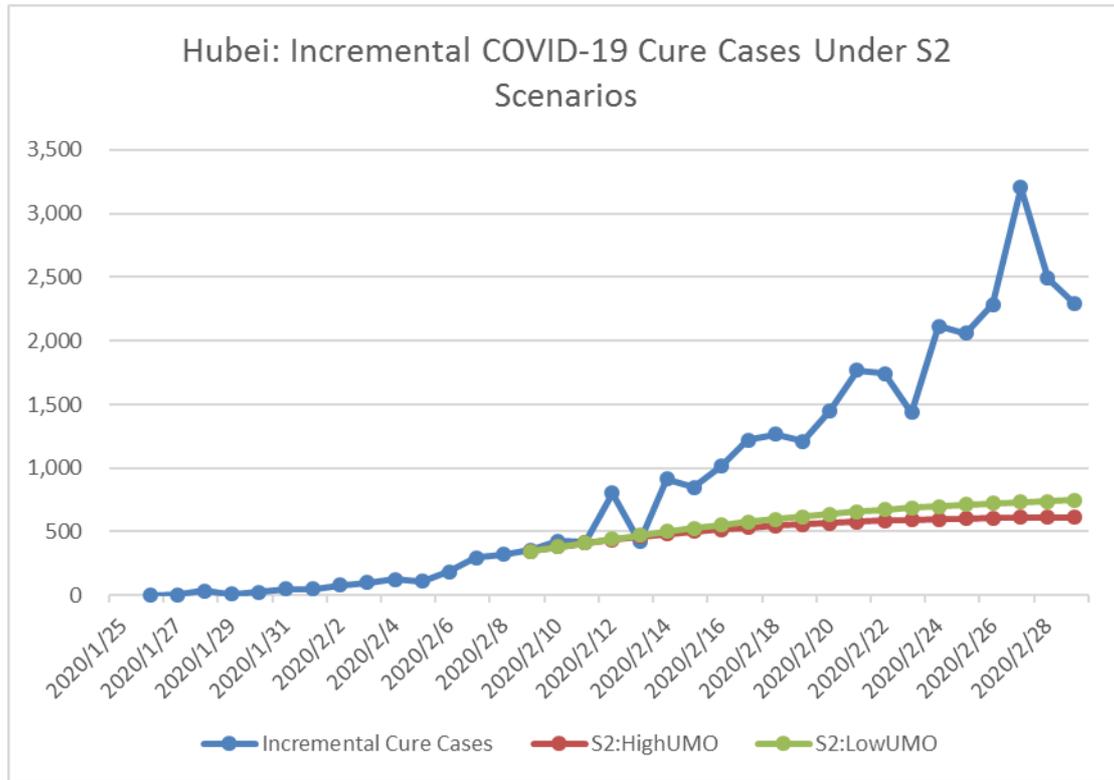

Figure 9. Cautiously Optimistic Scenario Forecasts of Cured Cases: Effect of Medical Assistance Teams Leads to Marked Increase in Cure Rates During February

## 4.4 Model Flexibility

As illustrated earlier, the TMM approach can forecast all states in the COVID-19 pandemic, including the intermediate states (medical observation, mild cases, severe cases, critical cases), and the terminal states (discharge from medical observation, cure, death). The model does not include asymptomatic cases. The inclusion of asymptomatic cases can be readily handled by adding one more intermediate state, although good data are not available for the asymptomatic population. Also, the major purpose of the model is to predict the active cases, i.e., patients who are receiving treatment in medical facilities, and the patient distribution of severe/critical cases to better manage medical resources. Thus, adding an additional asymptomatic state not only unnecessarily complicates the model, but also compromises its robustness.

In another paper (Zheng et. al 2020), we drew from experience in Hubei and other provinces of China during the early stages of the COVID-19 outbreak and improved our model to incorporate different levels of preventive policy effectiveness and made forecasts for Italy, South Korea, Iran. Later, we applied the model to forecast the COVID-19 progression in other countries, including Spain, France, Germany, and United States. Most of these models performed reasonably well, and the forecasts are included in Appendices I & II. We also estimate the preventive policy effectiveness parameter in Appendix III.

## 5. Supporting COVID-19 Medical Management Decisions

In the very early stages of the COVID-19 outbreak, the daily fatality rate of severe (including critical) cases was extremely high, around 8%. On Chinese New Year's Eve, January 24th, the first medical assistance team was dispatched from Shanghai to Wuhan to help the local medical staff. On February 10th, the TMM forecast was adopted by the first Shanghai medical assistance team (led by Dr. Zheng, one of this paper's authors) in Wuhan's Jinyintan Hospital, the first designated hospital to take COVID-19 patients in the world. The forecast has been used in preparing medical staff, ICU beds, ventilators, and other critical care medical resources by central and provincial health commissions and local centers of disease control. For example, on February 14th, we published an article[16] indicating that under the cautiously optimal scenario, medical staff needed for taking care of severe and critical patients could reach 40,000-45,000. Soon after this forecast, more medical assistance teams were dispatched from all over China to Wuhan and other Hubei province cities, reaching more than 42,000 medical staff by the beginning of March. On February 15th, we forecasted the "back-to-normal" date most likely to be mid-April[17], but due to the extraordinary efforts of these medical assistance teams, the lock-down in Wuhan was lifted on April 8th (and Dr. Zheng was able to return to Shanghai after fighting COVID-19 for 67 days in the epicenter of Wuhan).

During January/February 2020, two hospitals with 1,900 beds were built within two weeks to accept severe and critical patients. As time went on, medical staff became more experienced treating COVID-19. All these measures helped to reduce the daily severe case fatality rate to a very low level of 0.5%, an almost 94% drop from the very early stages, illustrated in Figure 10. The model forecasts in Figure 8 use the higher fatality rate observed on 2/8, resulting in higher death toll forecasts.

Moreover, starting from 2/5, 16 Fangcang shelter hospitals were put into use, accepting more than 12,000 non-severe patients in Wuhan. These shelter hospitals were converted from stadiums, shopping malls, convention centers, etc. This measure moved the treatment window earlier and placed patients with mild symptoms in these makeshift hospitals to receive proper medical treatment rather than self-quarantining at home, where there would be significantly heightened risk of family transmission and community transmission. The daily cure rate of non-severe cases increased dramatically from the level of 1% in late January to close to 10% in late February, illustrated in Figure 11. The model forecasts in Figure 9 use the much lower cure rate observed on 2/8, resulting in the significantly lower cure forecasts.

Finally, as a result of all these timely measures, the peak of actual active cases was moved earlier to February 16, instead of the date originally estimated as between 3/1-3/7, as illustrated in Figure 12.

---

[16] http://chenjian.blog.caixin.com/archives/221560
[17] http://chenjian.blog.caixin.com/archives/221630

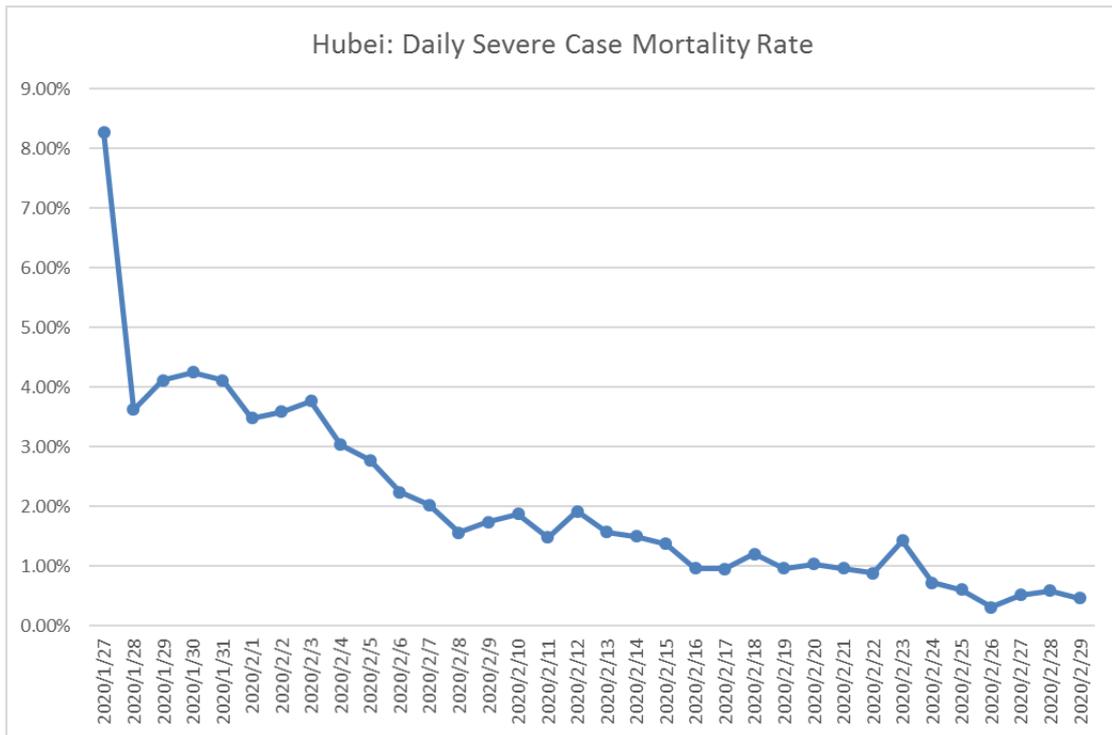

Figure 10. Mortality Rate Continual Drop (Nonstationarity Affects Model Forecasts)

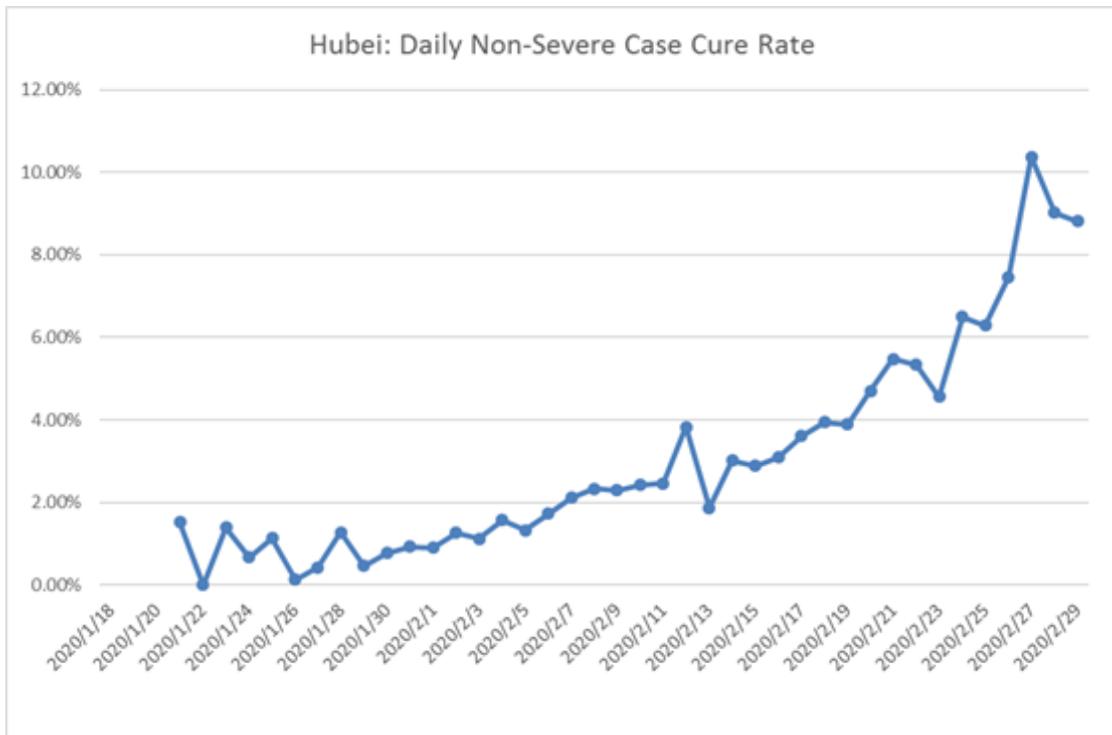

Figure 11. Cure Rate Continual Improvement (Nonstationarity Affects Model Forecasts)

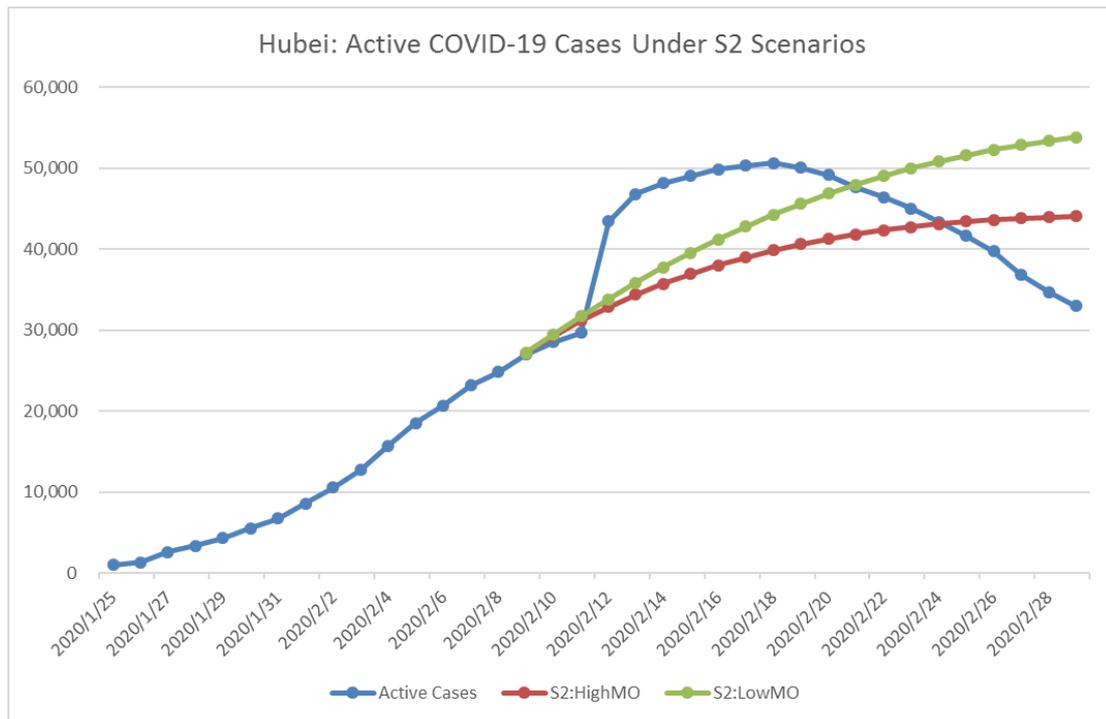

Figure 12. Actual Peak of Active Cases in mid-February Two Weeks Earlier than Model Forecasts Due to Improvement in Medical Treatment (Also Reflected in Figures 9 and 10)

## 6. Conclusions, Ongoing Research, and Key Takeaways

We introduced a new discrete-time Markov chain (DTMC) transition matrix model (TMM) for modeling epidemic outbreaks that directly incorporates stochastic behavior. Parameter estimation for the model is straightforward, so we applied the model using COVID-19 data from Hubei province, for which it provided reasonably accurate forecasts, with sensitivity analysis illustrating its robustness properties in terms of far less sensitivity to parameter misspecification than traditional epidemiological compartmental models. As a result, the model has been adopted by the first Shanghai assistance medical team in Wuhan's Jinyintan Hospital, the first designated hospital to take COVID-19 patients in the world, and the forecasts have been used for preparing and allocating medical staff, ICU beds, ventilators, and other critical care medical resources and for supporting medical management decisions.

The proposed approach can forecast all the states in the COVID-19 pandemic, including the intermediate states (medical observation, mild cases, severe cases, critical cases), and the terminal states (discharged from medical observation, cured, deceased). As mentioned earlier in the model formulation, the model applied here to COVID-19 does not include asymptomatic cases, which can be readily handled by adding one more intermediate state, but was purposely excluded since the data for the asymptomatic population is not available. Also, the major purpose of the model is to predict active cases, i.e., patients who are receiving treatment in medical facilities, and the patient distribution of severe/critical cases to better manage medical resources.

Similar to the finance academic community where the literature on asset pricing is dominated by stochastic partial differential equation models for which stylized models are used to generated closed-form solution and big-picture insights, the prevalent modeling paradigm in the academic epidemiological research literature also values stylized models, in this case based on systems of deterministic ordinary differential equations. The first author has spent over two decades in the finance industry and implemented many models for investment decisions, and believes that the model proposed here follows the same vein of industry relying on more practically implementable models. The proposed approach is a preliminary attempt to advocate models that are tailored to the available data and anticipated usage of the model in decision making, whether it be strategic policy, supply chain planning, or hospital operations. Thus, the focus is on flexibility, ease of implementation, and robustness rather than theoretical elegance.

In terms of ongoing work, Zheng et al. (2020) draws upon the experience from Hubei and other provinces in China during the early stages of the COVID-19 outbreak to improve the TMM model by incorporating different levels of preventive policy efficiency. This led to forecasts for Italy, South Korea, and Iran, which were posted online[18] on March 9. The forecasts for Italy were channeled to an Italian cabinet member on the same day, indicating a very dire situation with forecast of more than 190,000 likely cases by April end with weak intervention efforts. The Italian government implemented a national "lock-down" policy on the next day.

The best COVID-19 forecast model probably does not exist, but some models are more useful than others, depending on the types of decision being supported. Traditional epidemiology models, like the SIR-type models rely heavily on estimated parameters such as R0 and could have very wide forecast ranges over a relatively short period, due to high sensitivity and wide confidence intervals of key parameters. They are useful for reference purposes but may have limited ability in supporting real-time decision making. TMM models are mainly driven by empirical probabilities, and do not require complicated estimation procedures. In our case study, they are shown to be more robust, flexible, and accurate, once the clinical experts are involved in the modeling process. They know more about fighting COVID-19 in the frontline. Their experience is of immeasurable value not only in saving lives directly, but also for building models that can help save lives through efficient allocation of medical resources.

In the eyes of a hammer, everything is a nail. But it's not true, so don't go with the "standard" models in the academic literature if they don't fit the use. Fit your best model to the data, not the other way around.

---

[18] http://chenjian.blog.caixin.com/archives/223401

# Appendix I. Forecasts of Italy, South Korea, and Iran on March 9, observed on April 18

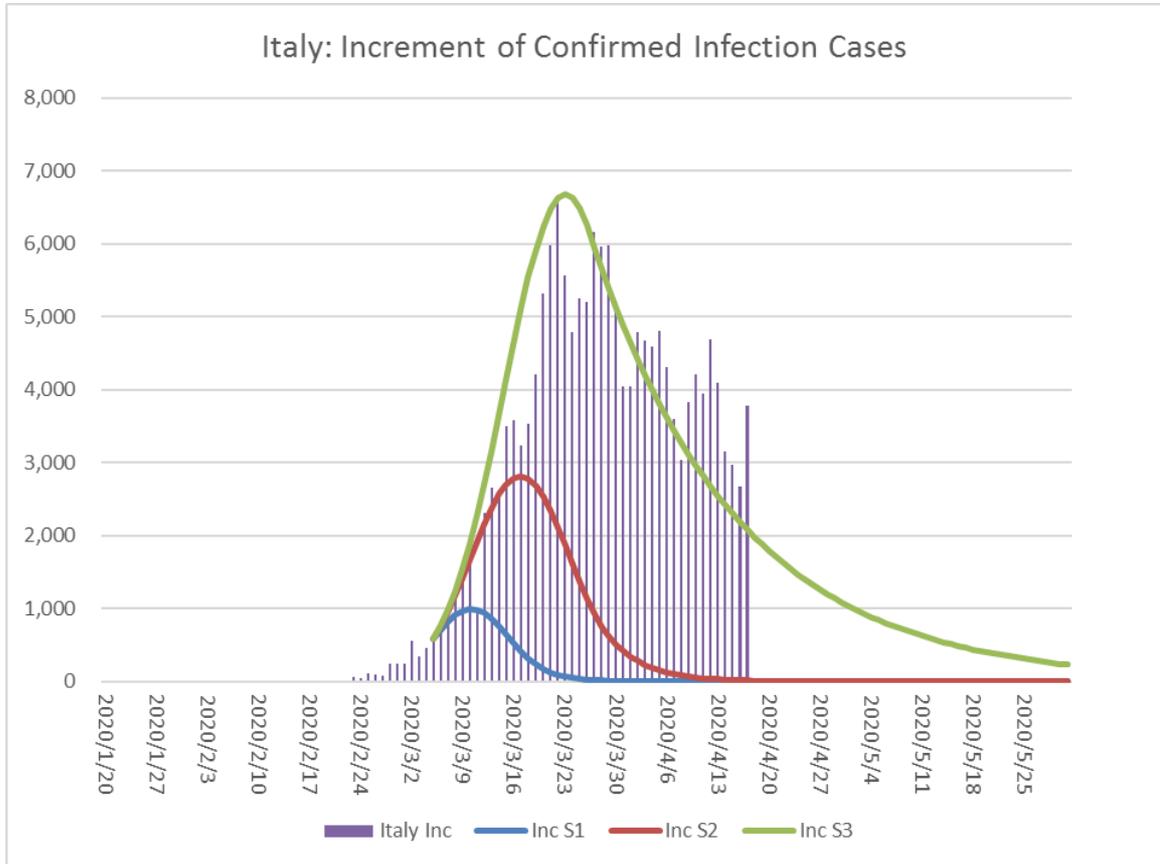

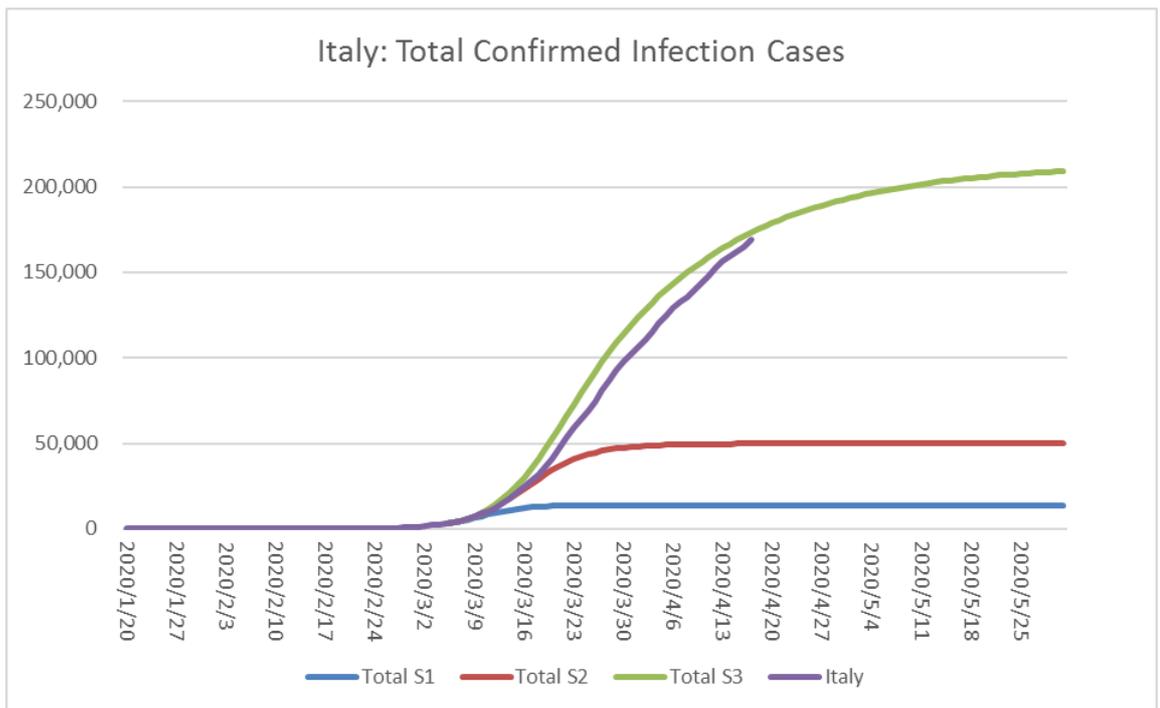

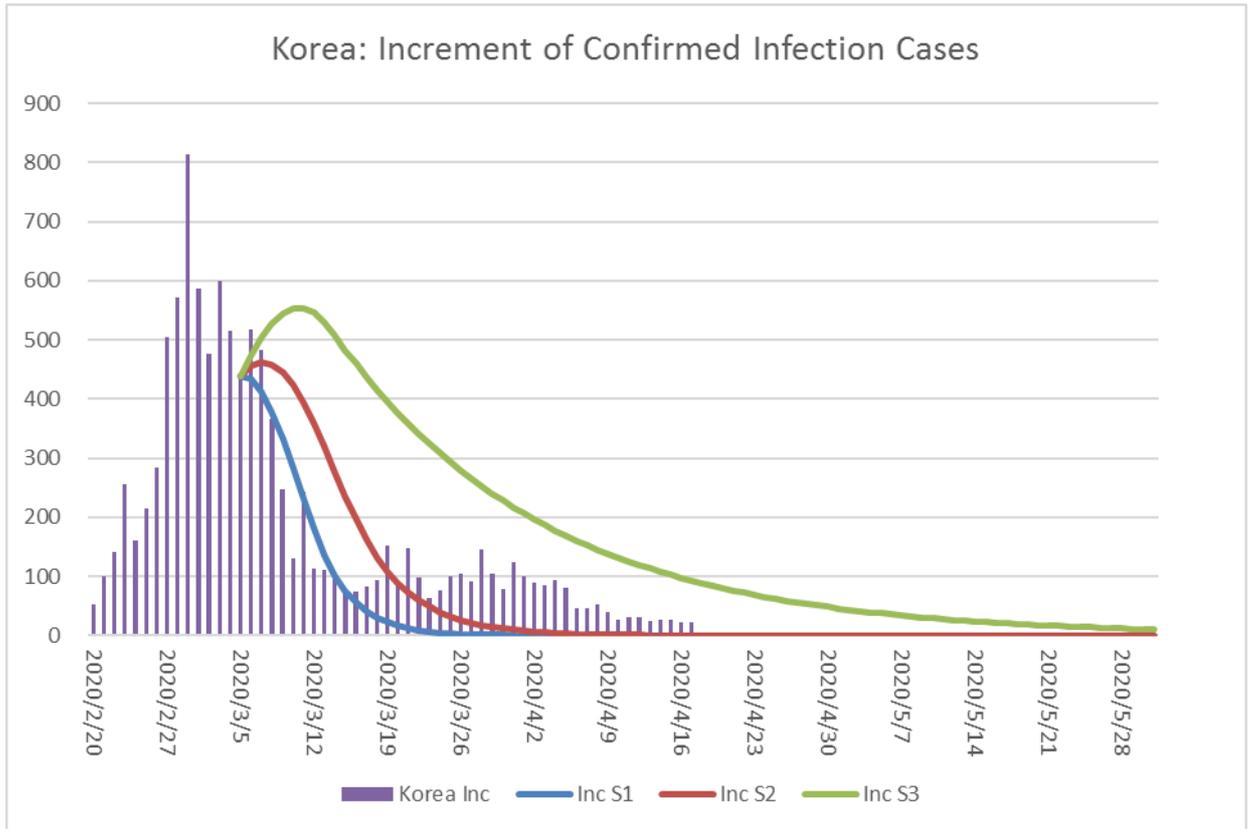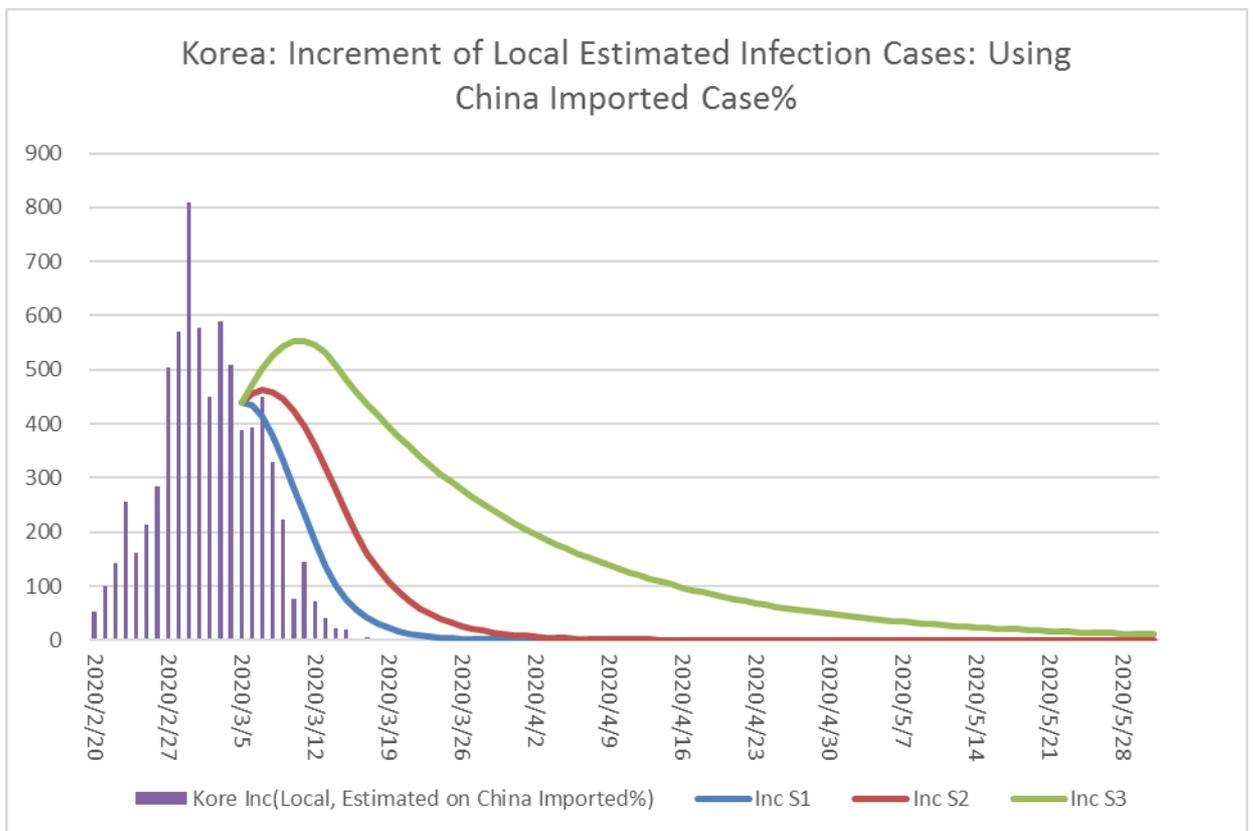

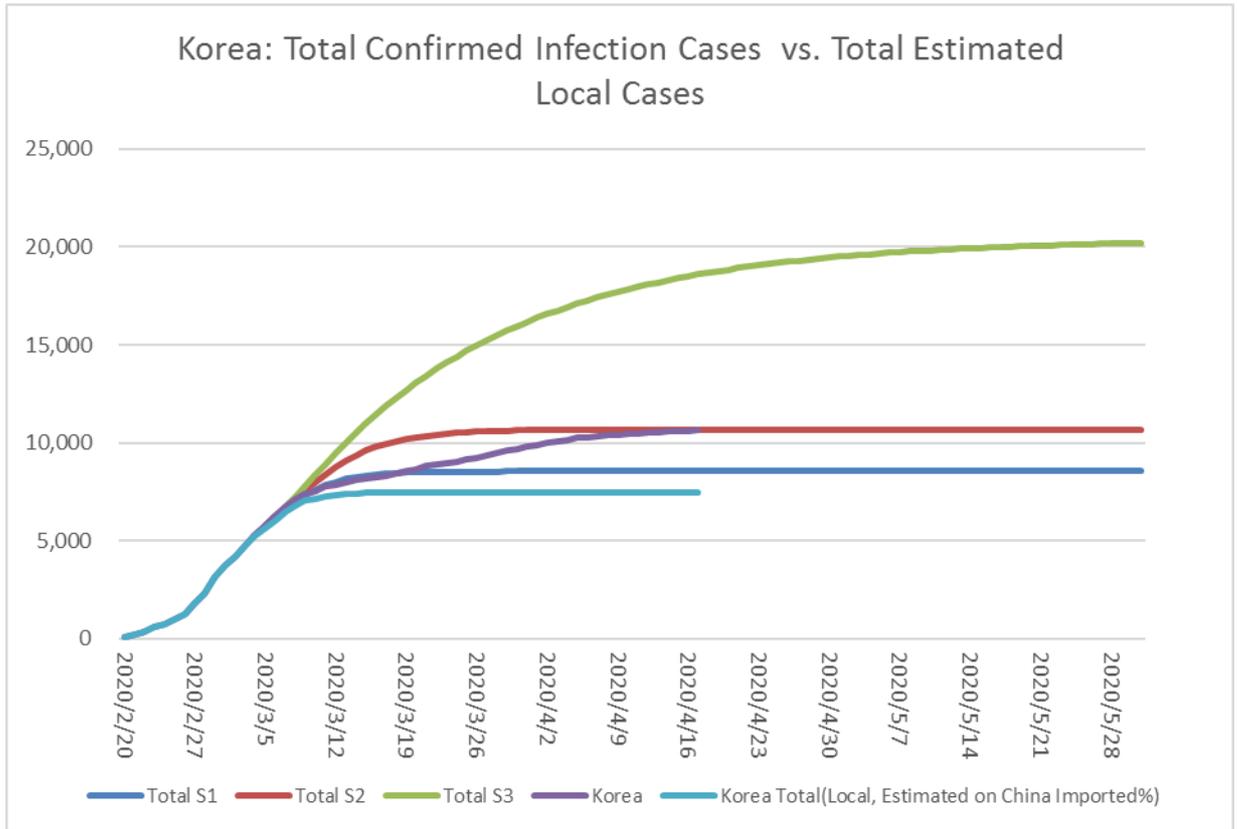

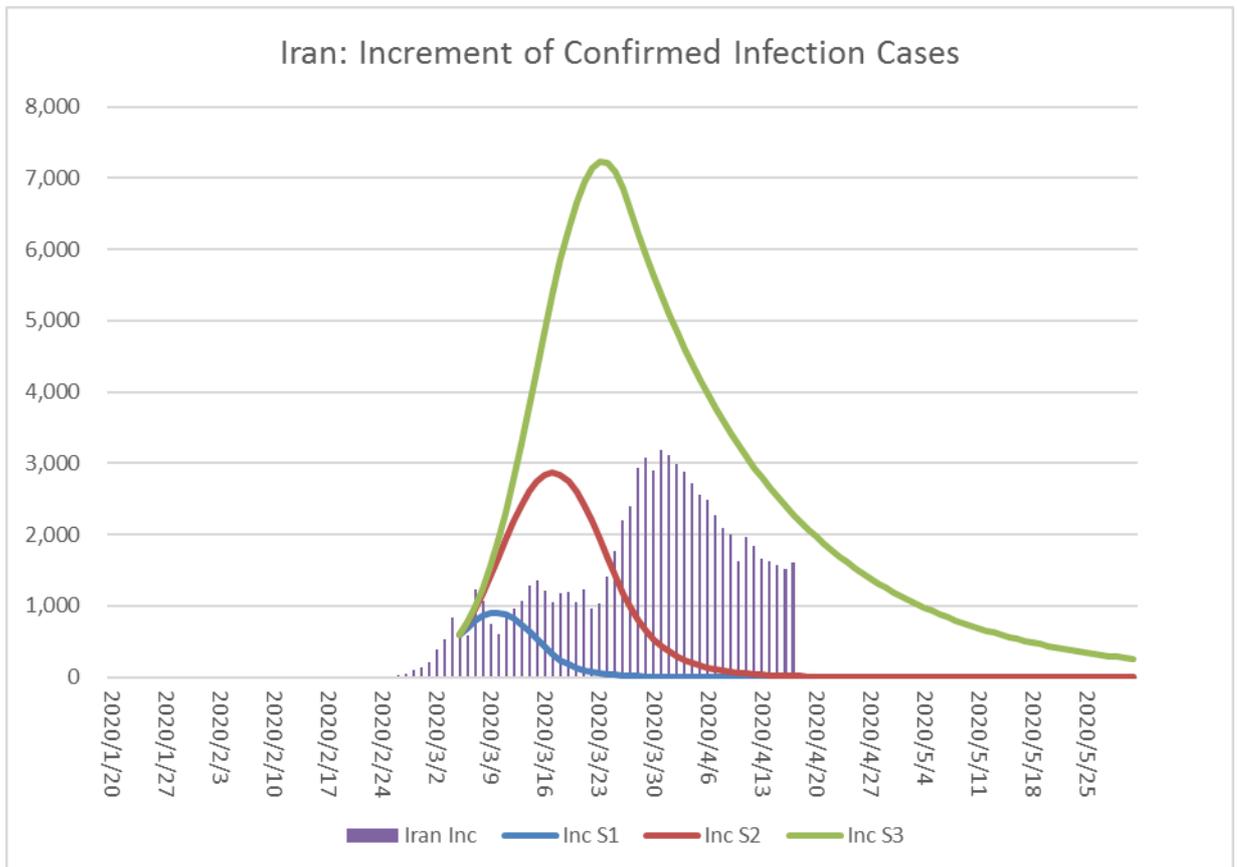

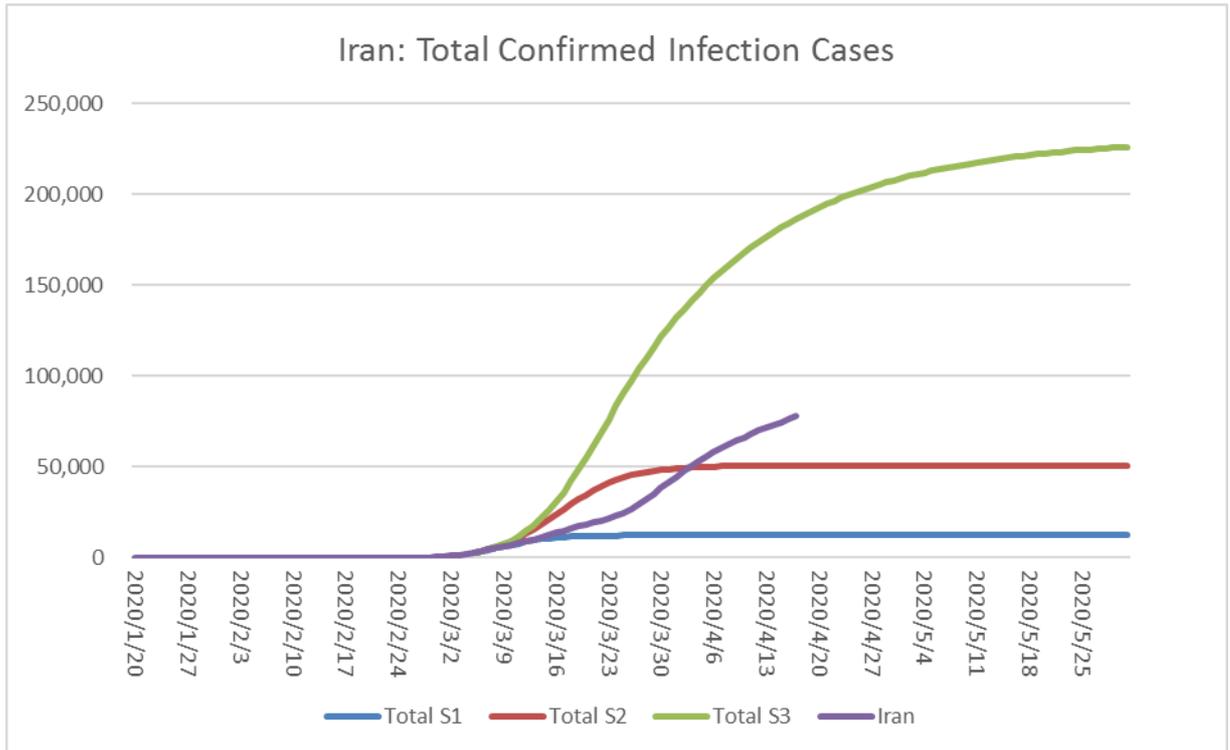

## Appendix II. Forecasts of Spain, France, Germany, USA on March 24, as Observed on April 18

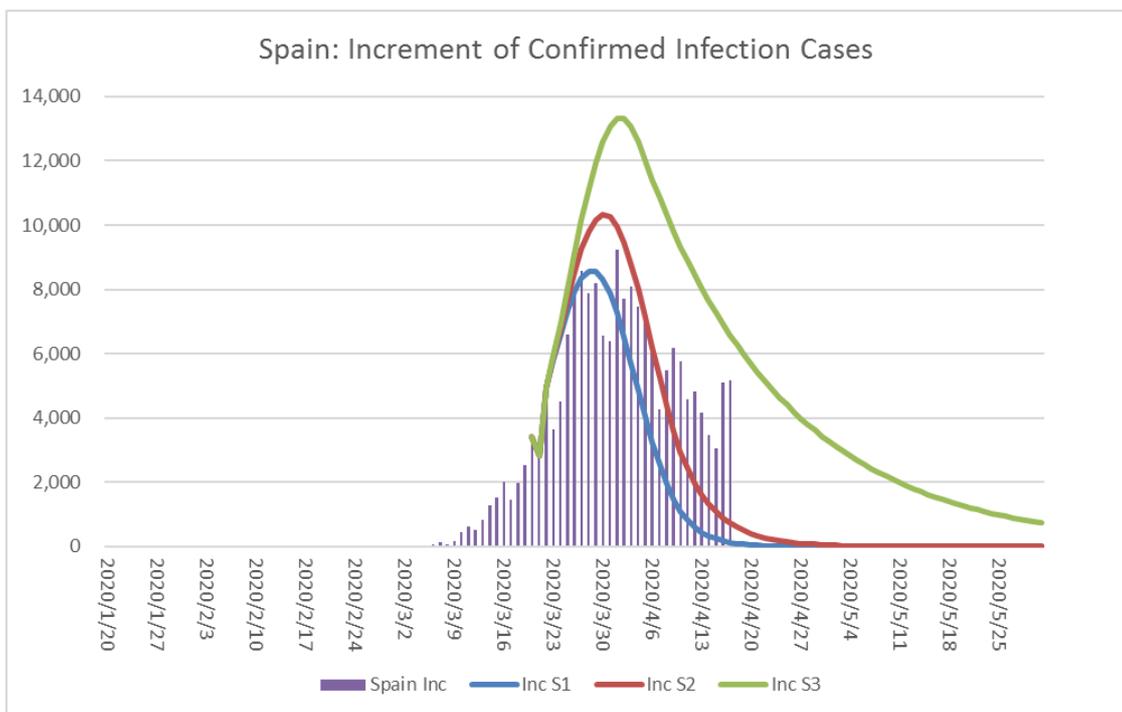

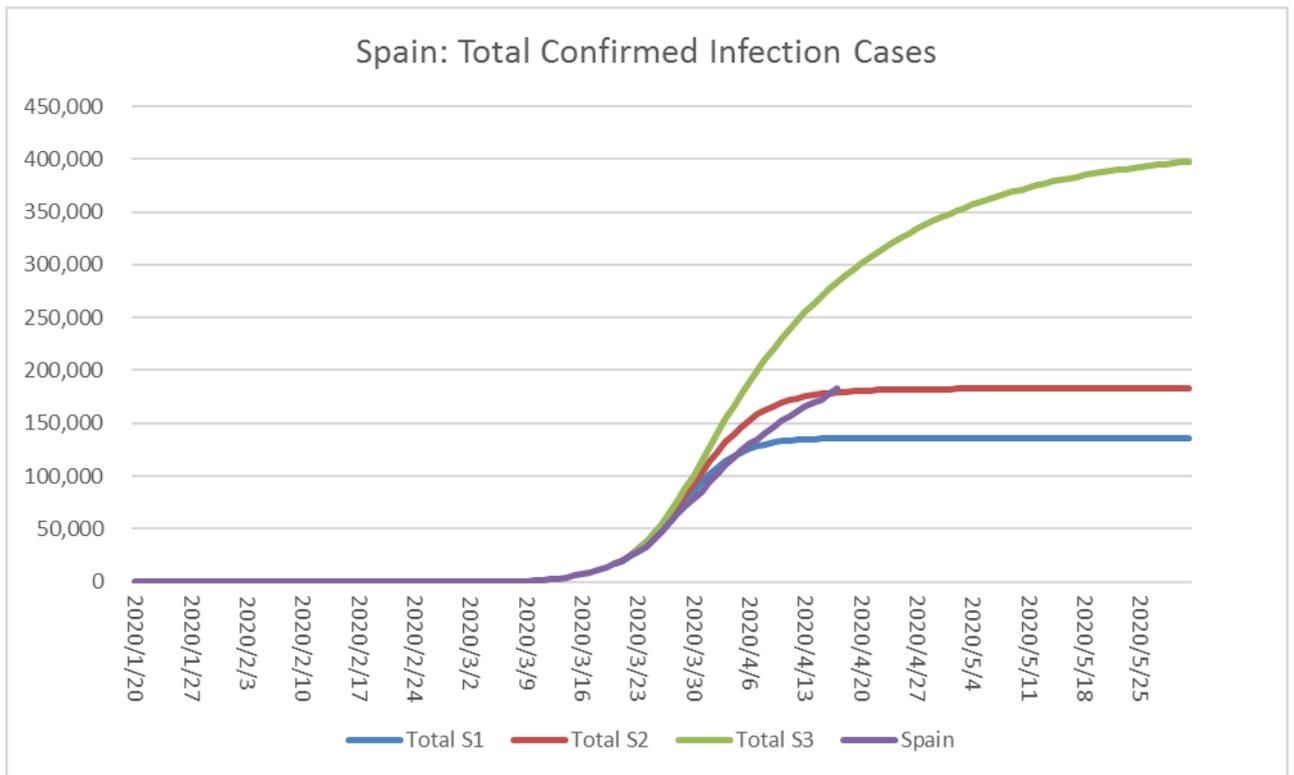

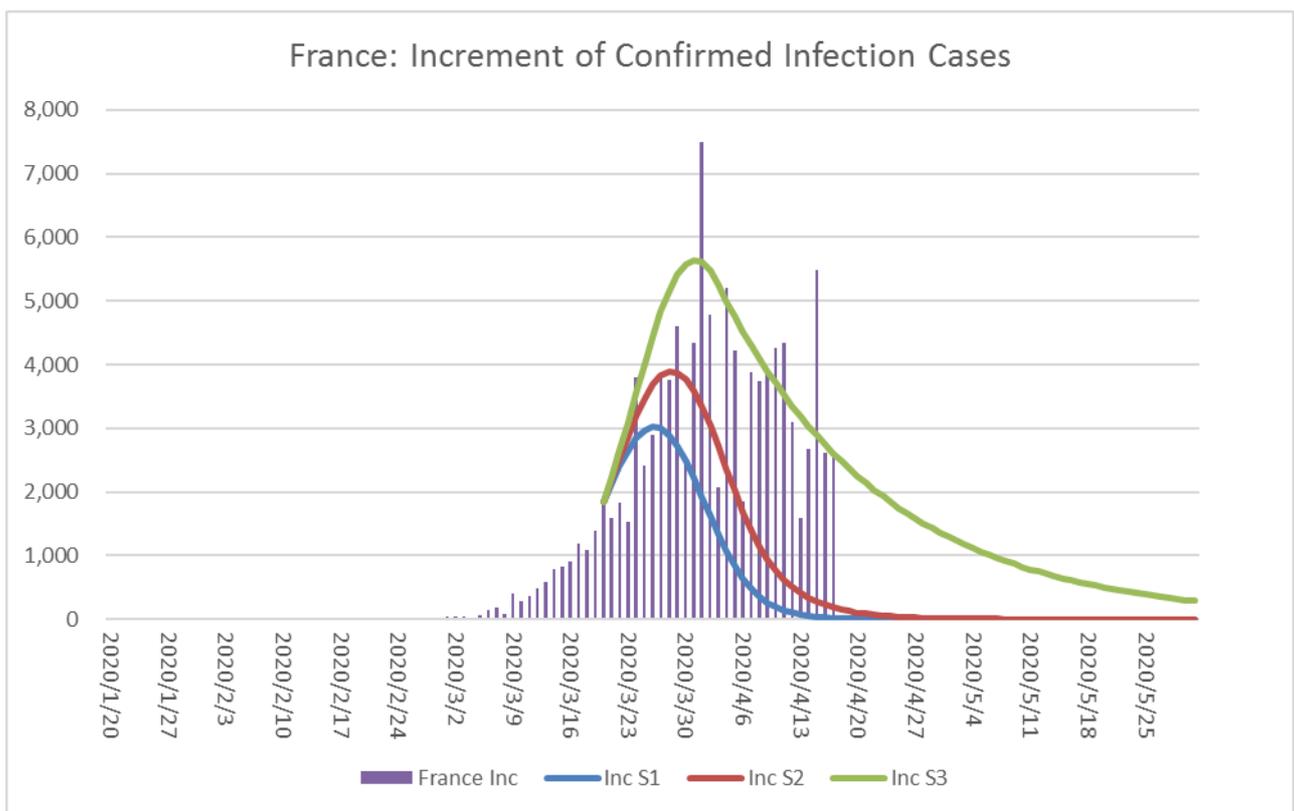

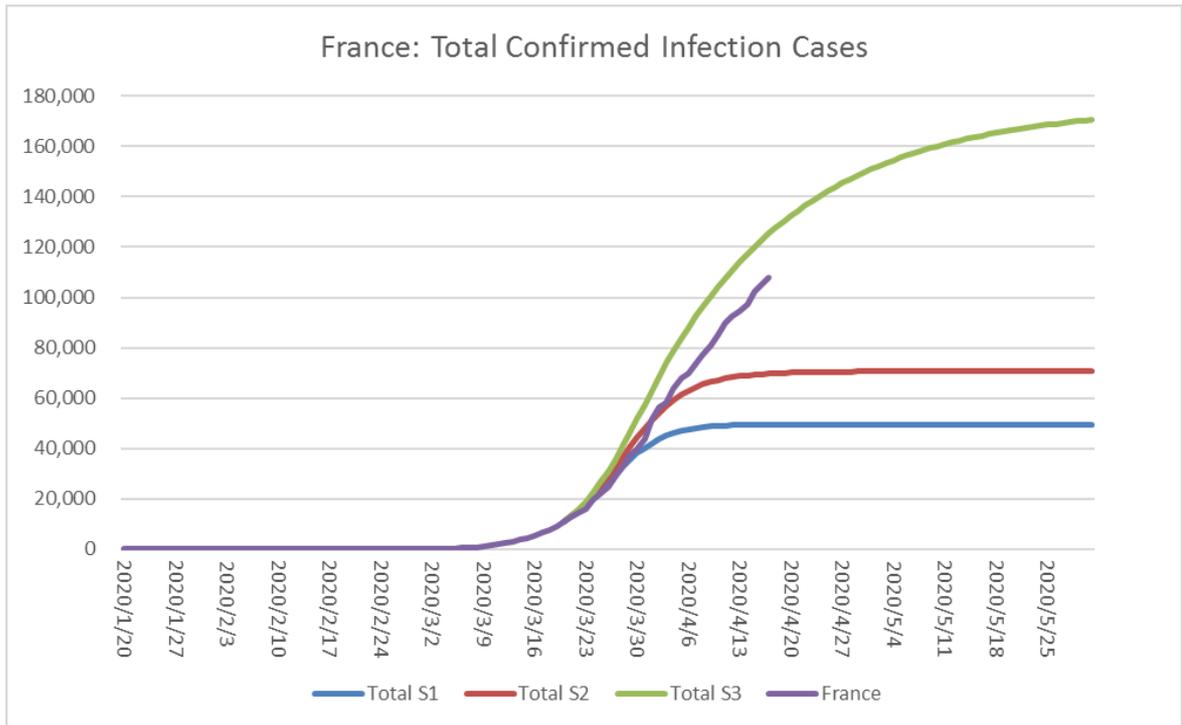

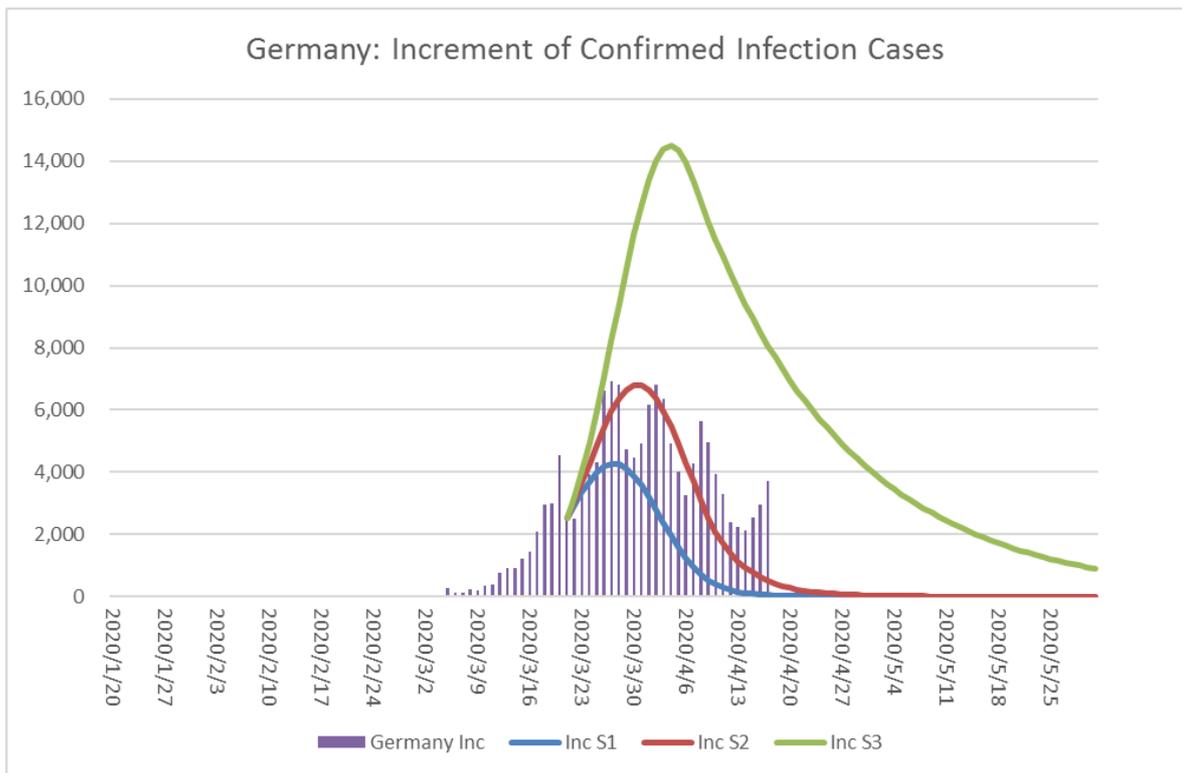

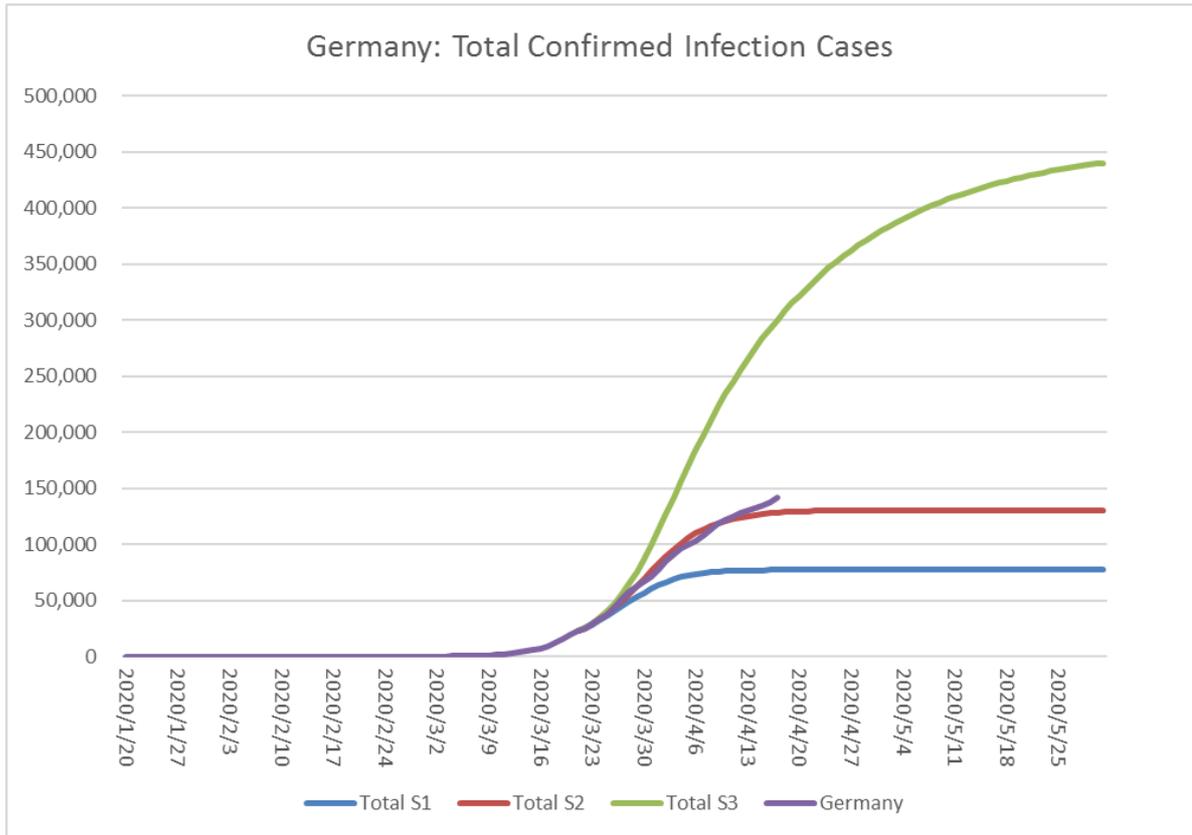

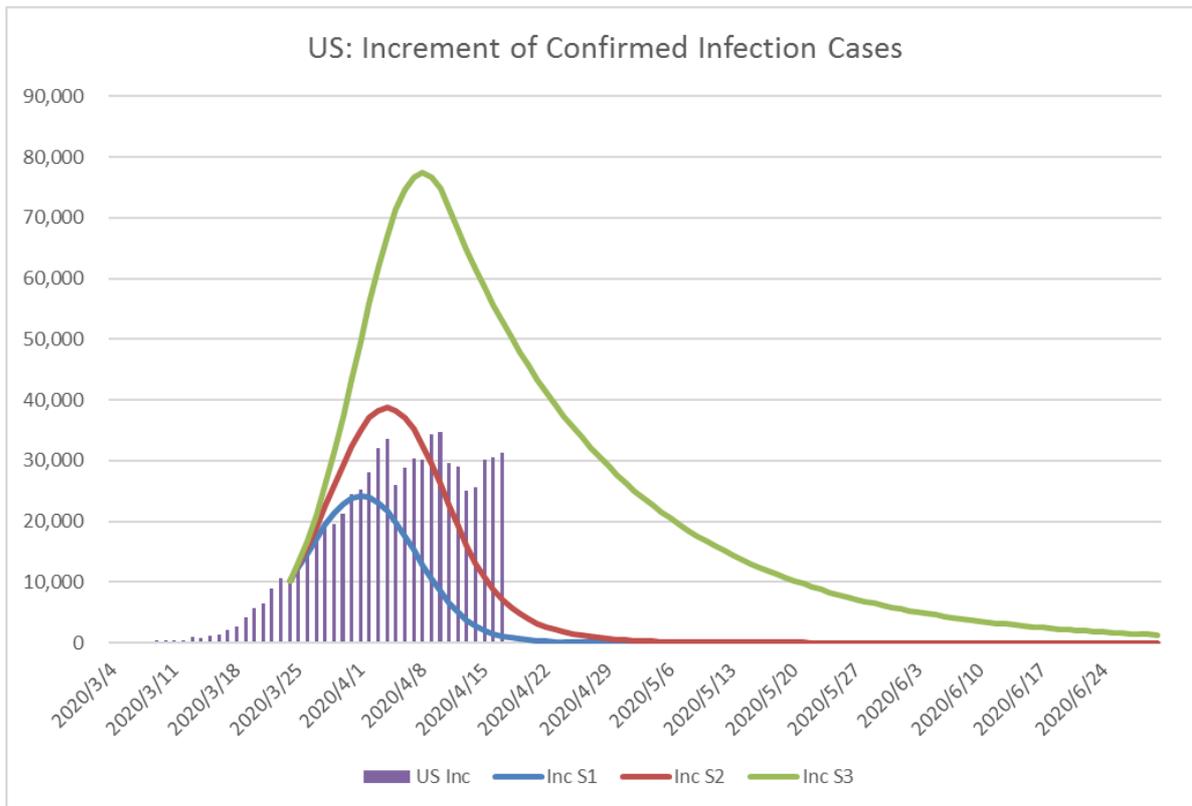

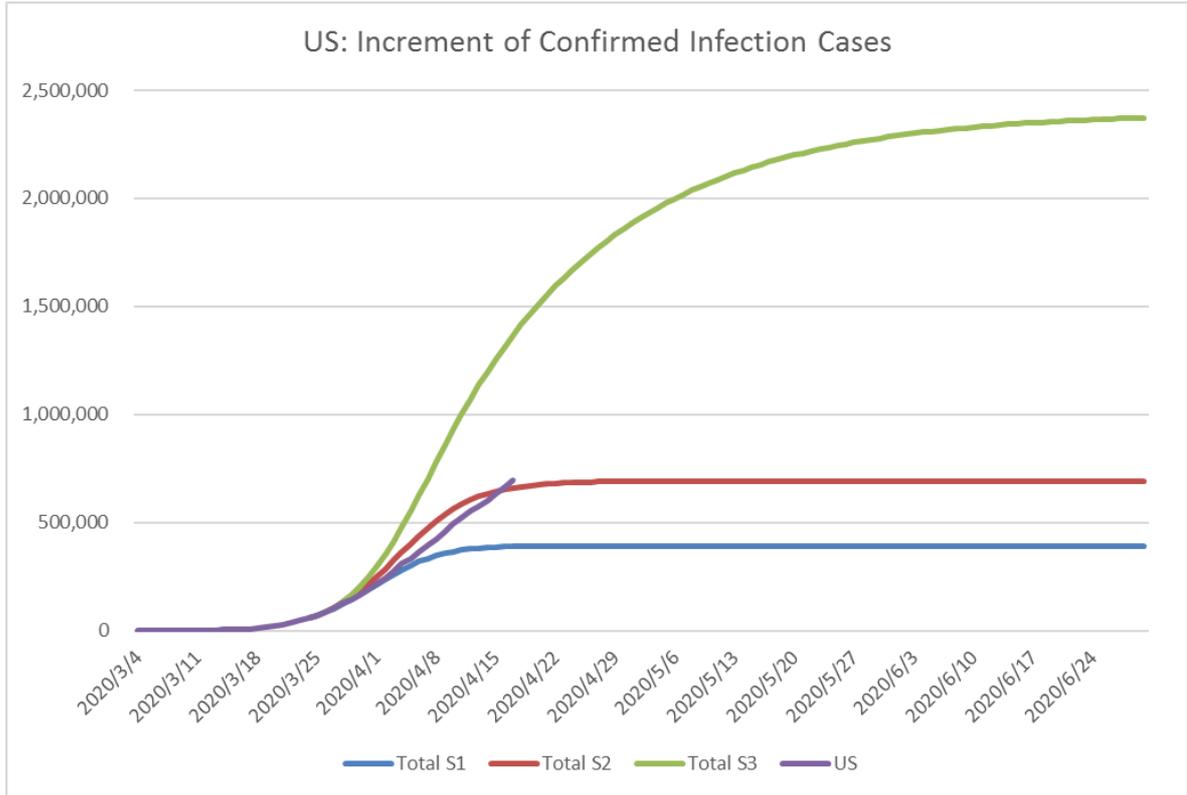

## Appendix III: Estimation of Daily New Cases Growth Rate Slope for Major Epidemic Countries/Areas

|  | Slope of Daily Inc% (10-Day MA) | $R^2$ |
|---|---|---|
| China-Ex. Hubei | -0.0387 | 0.8830 |
| China-Hubei | -0.0251 | 0.9369 |
| S. Korea | -0.0076 | 0.2992 |
| S. Korea (Revised) | -0.0352 | 0.7545 |
| Italy | -0.0086 | 0.8461 |
| Iran | -0.0097 | 0.6506 |
| Spain | -0.014 | 0.9297 |
| France | -0.0103 | 0.8512 |
| Germany | -0.0123 | 0.8816 |
| UK | -0.0079 | 0.7355 |
| USA | -0.0136 | 0.8389 |

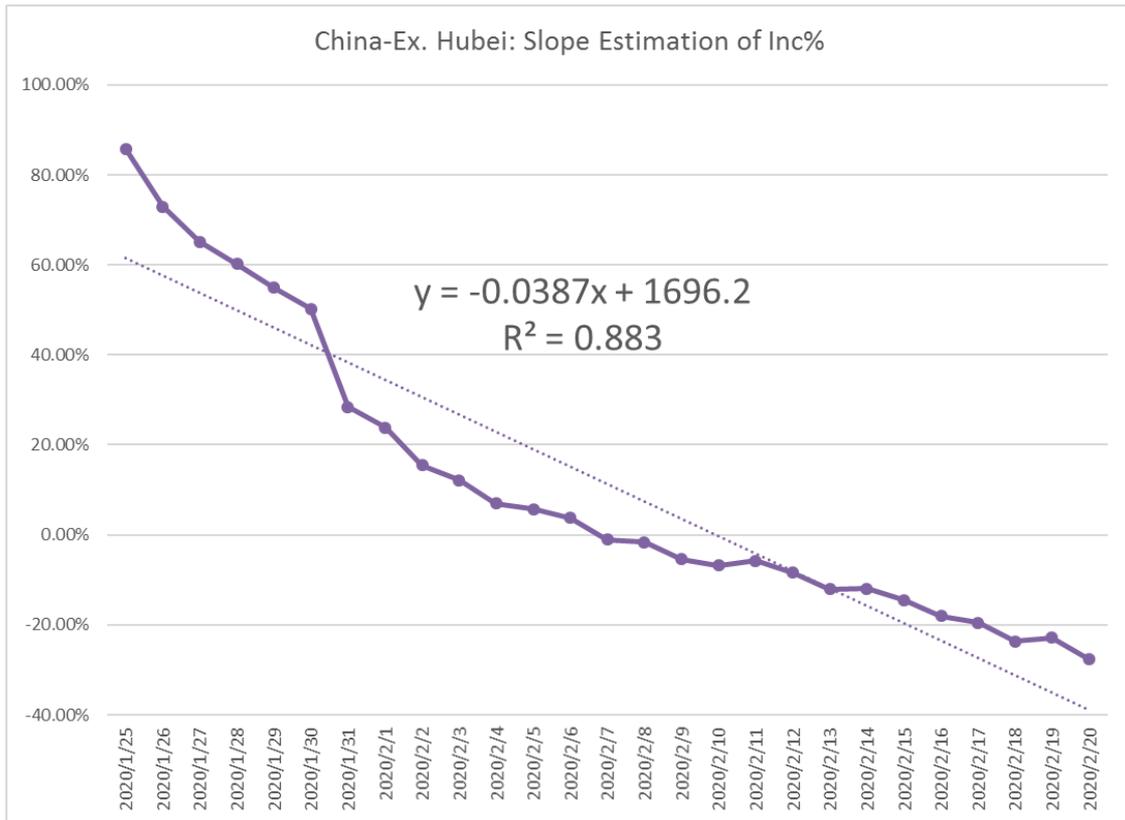

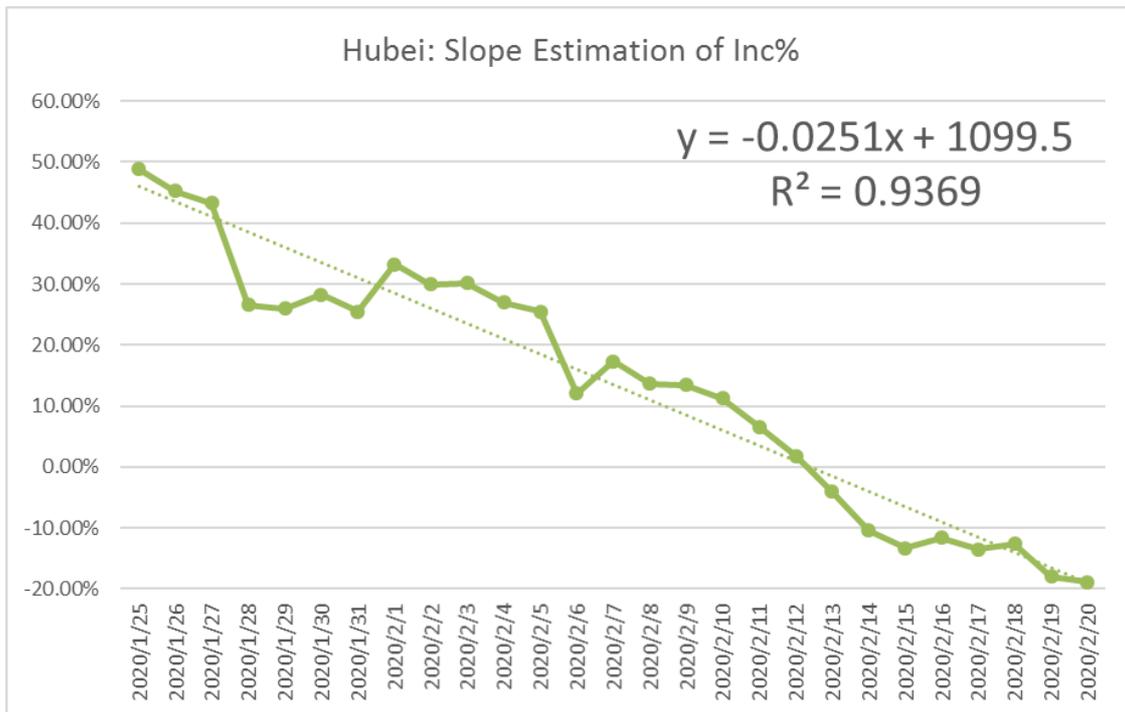

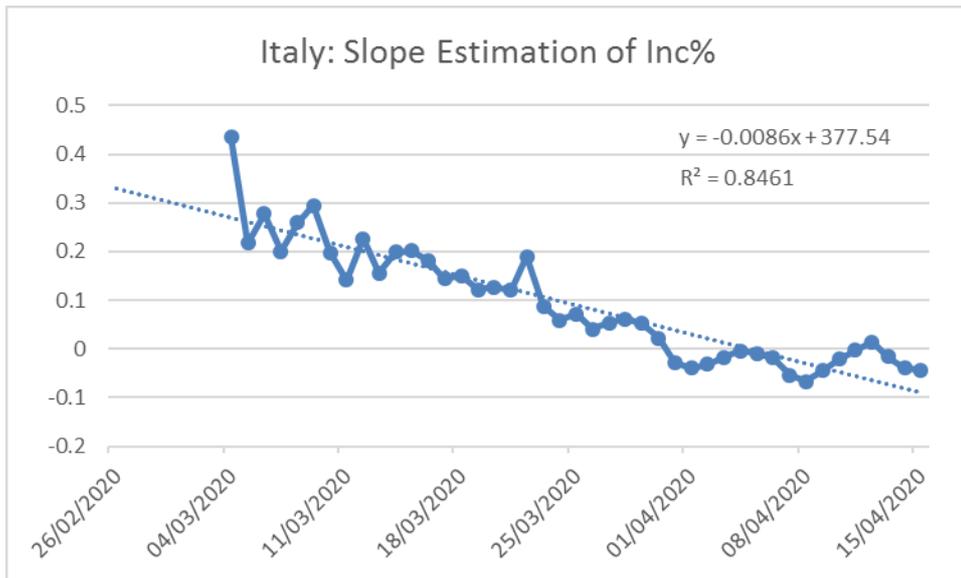
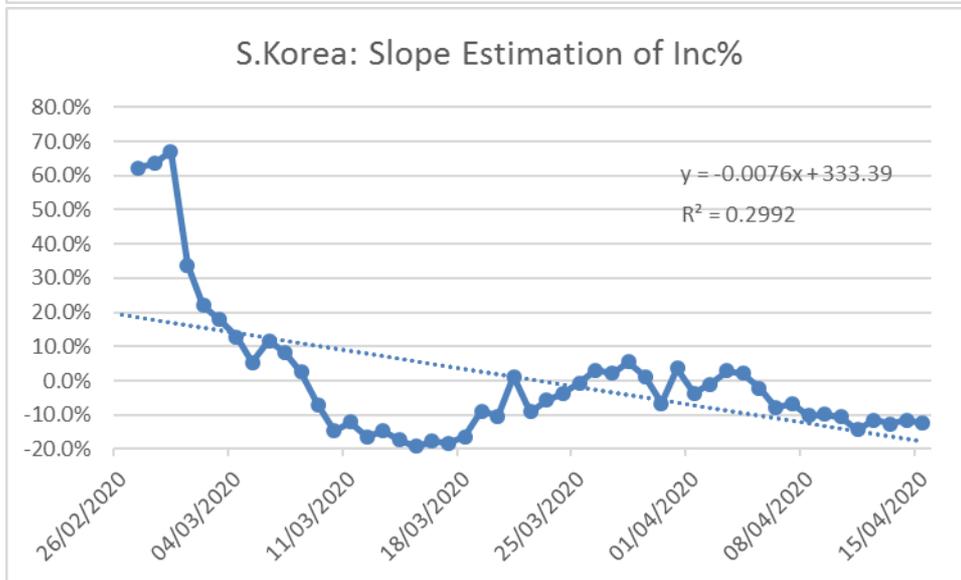
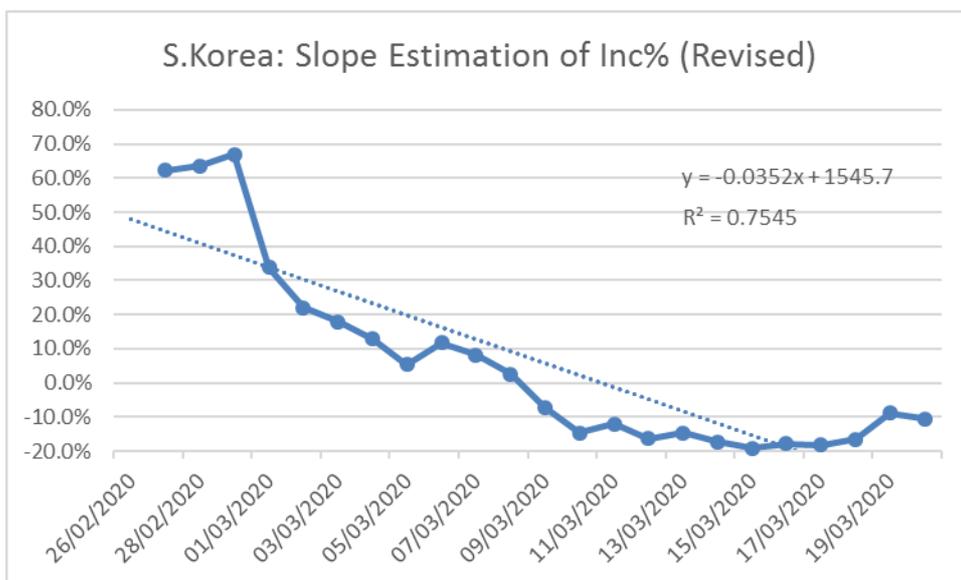

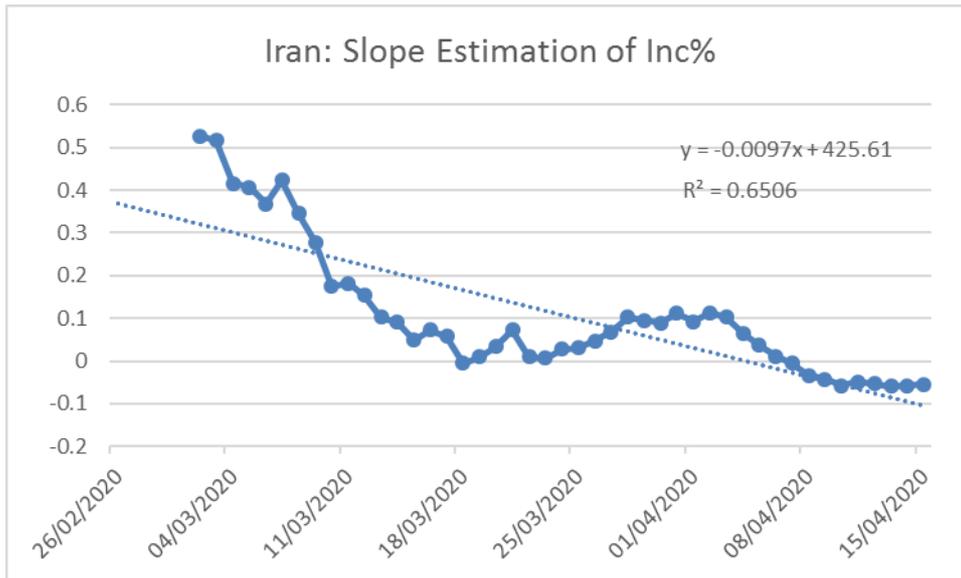

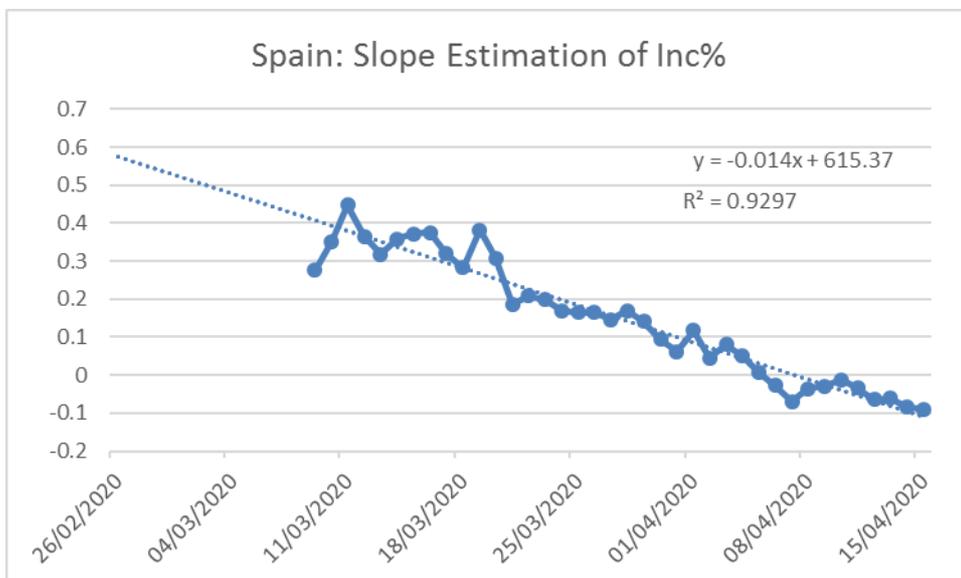

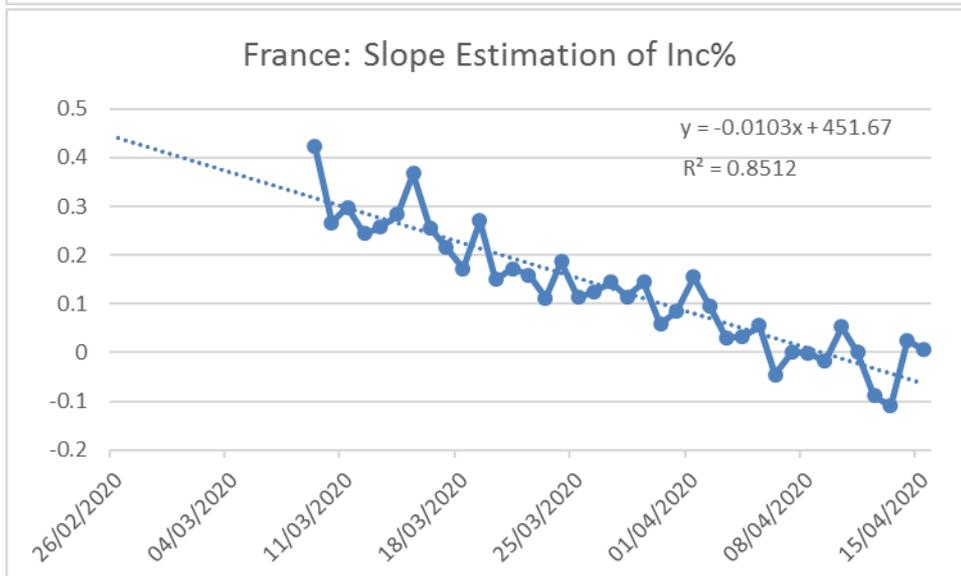

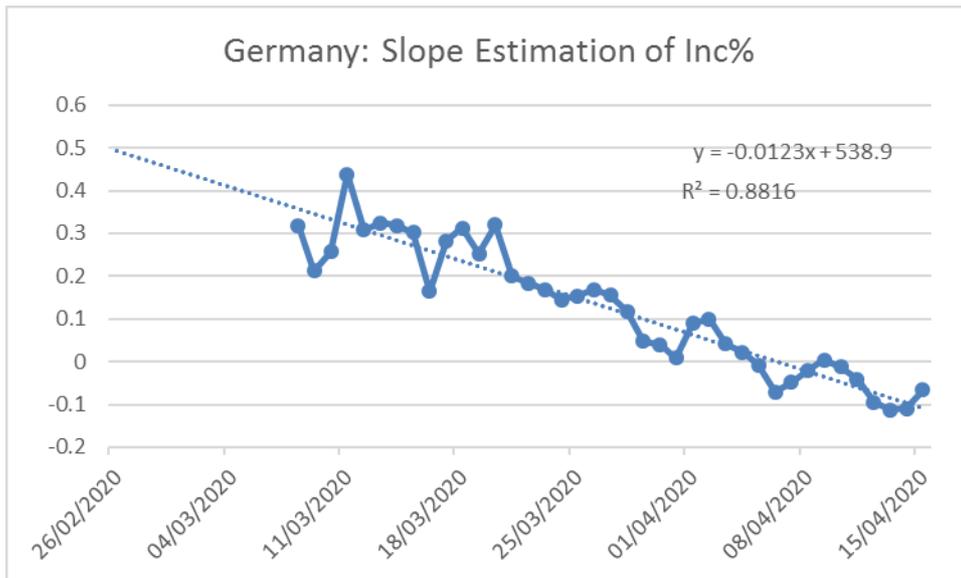
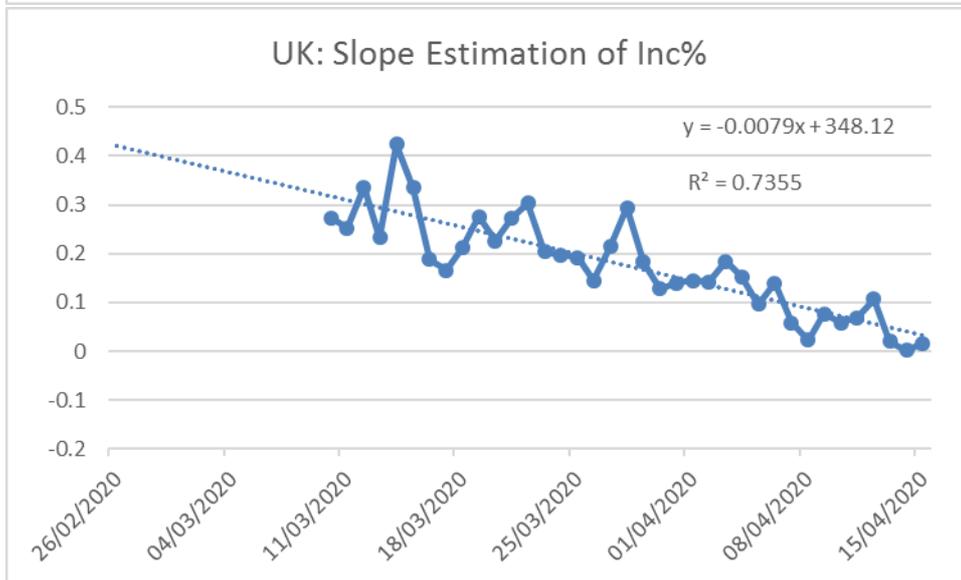
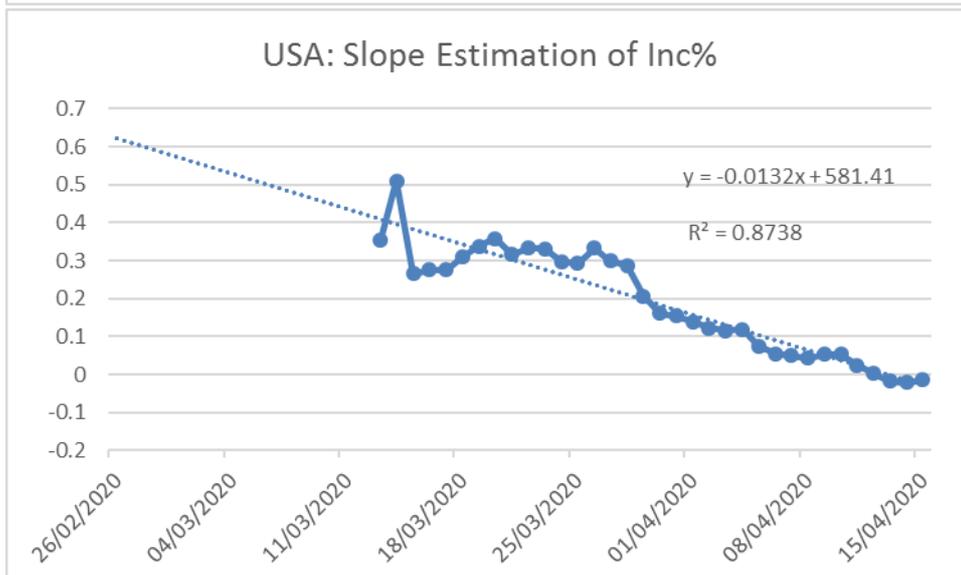